\documentclass{aa}
\usepackage{amsmath,amsfonts, amssymb}
\usepackage{graphicx,color}
\usepackage{txfonts}
\usepackage{longtable,booktabs}
\usepackage{xspace}
\usepackage{lscape}
\usepackage{tablefootnote}

\newcommand{\hi}{\textsc{H$\,$i}\xspace}
\newcommand{\htwo}{\ensuremath{\mathrm{H}_2}\xspace}
\newcommand{\numunit}[2]{\mbox{\ensuremath{#1\,#2}\xspace}}
\newcommand{\mstar}{\ensuremath{\text{M}_\star}\xspace}
\newcommand{\mhi}{\ensuremath{\text{M}_\hi}\xspace}
\newcommand{\kms}{\ensuremath{\text{km}\,\text{s}^{-1}}\xspace}
\newcommand{\cmt}{\ensuremath{\text{cm}^{-2}\xspace}}
\newcommand{\msun}{\ensuremath{\text{M}_\odot}\xspace}
\newcommand{\mk}{MeerKAT\xspace}

\newcommand{\mhz}{\text{MHz}\xspace}
\newcommand{\mpc}{\text{Mpc}\xspace}
\newcommand{\kpc}{\text{kpc}\xspace}

\newcommand{\sofia}{\texttt{SoFiA-2}\xspace}

\newcommand{\nhi}{\ensuremath{\mathrm{N}_{\mathrm{H}\,\textsc{i}}}\xspace}
\newcommand{\gal}{NGC 5068\xspace}

\newcommand{\secref}[1]{Sec.~\ref{#1}}
\newcommand{\figref}[1]{Fig.~\ref{#1}}
\newcommand{\tabref}[1]{Table~\ref{#1}}

\graphicspath{{images/}}

\begin{document}

   \title{Possible origins of anomalous H$\,$\textsc{i} gas around MHONGOOSE galaxy, NGC 5068}
   \titlerunning{NGC 5068}
   \authorrunning{Healy et al.}

   \subtitle{}

   \author{J. Healy,\inst{1} \thanks{healy AT astron.nl}, 
   W.J.G. de Blok,\inst{1,2,3}, 
   F.M. Maccagni\inst{1,4},
   P. Amram\inst{5},
   L. Chemin\inst{6},
   F. Combes\inst{7,8},
   B.W. Holwerda\inst{9},
   P. Kamphuis\inst{10},
   D.J. Pisano\inst{3},
   E. Schinnerer\inst{11},
   K. Spekkens\inst{12,13}, 
   L. Verdes-Montenegro\inst{14},
   F. Walter\inst{11},
   E.A.K. Adams\inst{1,2},
   B.K. Gibson\inst{15},
   D. Kleiner\inst{1,4},
   S. Veronese\inst{1,2},
   N. Zabel\inst{3},
   J. English\inst{16}
   \and
   C. Carignan\inst{3,17,18}
   } 

  \institute{$^{1}$Netherlands Institute for Radio Astronomy (ASTRON), Oude Hoogeveensedijk 4, 7991 PD Dwingeloo, The Netherlands\\
  $^2$Kapteyn Astronomical Institute, University of Groningen, PO Box 800, 9700 AV Groningen, The Netherlands,\\
  $^3$Department of Astronomy, University of Cape Town, Private Bag X3, 7701 Rondebosch, South Africa,\\
  $^4$INAF -- Osservatorio Astronomico di Cagliari, via della Scienza 5, 09047, Selargius (CA), Italy,\\
  $^5$Aix-Marseille Univ., CNRS, CNES, LAM, 38 rue Frédéric Joliot Curie, 13338 Marseille, France\\
  $^6$Instituto de Astrofisica, Facultad de Ciencias Exactas,Universidad Andres Bello, Chile \\
  $^7$LERMA, Observatoire de Paris, PSL research Université, CNRS, Sorbonne Universit\'e, 75104, Paris, France, \\
  $^8$Collège de France, 11 Place Marcelin Berthelot, 75005, Paris, France,\\
  $^9$University of Louisville, Department of Physics and Astronomy, 102 Natural Science Building, 40292 KY Louisville, USA,\\
  $^{10}$Ruhr University Bochum, Faculty of Physics and Astronomy, Astronomical Institute (AIRUB), 44780 Bochum, Germany,\\
  $^{11}$Max-Planck-Institut f\"{u}r Astronomie, K\"{o}nigstuhl 17, D-69117, Heidelberg, Germany,\\
  $^{12}$Department of Physics and Space Science, Royal Military College of Canada P.O. Box 17000, Station Forces Kingston, ON K7K 7B4, Canada,\\
  $^{13}$Department of Physics, Engineering Physics and Astronomy, Queen’s University, Kingston, ON K7L 3N6, Canada,\\
  $^{14}$Instituto de Astrofísica de Andalucía (CSIC), Glorieta de la Astronomia s/n, 18008 Granada, Spain,\\
  $^{15}$E.A. Milne Centre for Astrophysics, University of Hull, HU6 7RX, United Kingdom,\\
  $^{16}$Department of Physics and Astronomy, University of Manitoba, Winnipeg, Manitoba R3T 2N2, Canada,\\
  $^{17}$ D\'{e}partement de physique, Universit\'{e} de Montr\'{e}al,  Complexe des sciences MIL, 1375 Avenue Th\'{e}r\`{e}se-Lavoie-Roux Montr\'{e}al, Qc, Canada H2V 0B3,\\
  $^{18}$Laboratoire de Physique et de Chimie de l'Environnement, Observatoire d'Astrophysique de l'Universit\'{e}  Ouaga I Pr Joseph Ki-Zerbo (ODAUO), BP 7021, Ouaga 03, Burkina Faso 
}
   \date{12 February 2024}

\abstract{
The existing reservoirs of neutral atomic hydrogen gas (\hi) in galaxies are insufficient to have maintained the observed levels of star formation without some kind of replenishment. {This refuelling of the \hi reservoirs} is likely to occur at column densities an order of magnitude lower than previous observational limits (\numunit{\mathrm{N}_{\hi,\, limit} \sim 10^{19}}{\cmt} at 30\arcsec\ resolution over a linewidth of \numunit{20}{\kms}). In this paper, we present recent deep \hi observations of NGC 5068, a nearby isolated star-forming galaxy observed by MeerKAT as part of the MHONGOOSE survey. With these new data, we are able to detect low column density \hi around \gal with a $3\sigma$ detection limit of \numunit{\nhi = 6.4 \times 10^{17}}{\cmt} at 90\arcsec\ resolution over a \numunit{20}{\kms} linewidth. The high sensitivity and resolution of the MeerKAT data reveal a complex morphology of the \hi in this galaxy -- a regularly rotating inner disk coincident with the main star-forming disk of the galaxy, a warped outer disk of low column density gas (\numunit{\nhi < 9\times 10^{19}}{\cmt}), in addition to clumps of gas on the north west side of the galaxy. We employ a simple two disk model that describe the inner and outer disks, and are able to identify anomalous gas that deviates from the rotation of the main galaxy. The morphology and the kinematics of the anomalous gas suggest a possible extra-galactic origin. We explore a number of possible origin scenarios that may explain the anomalous gas, and conclude that fresh accretion is the most likely scenario.
}

  \maketitle

    
\section{Introduction}
    \label{sec:intro}
    
How galaxies have been able to sustain the observed levels of star formation is one of the key open questions in galaxy evolution. In order for the current observed rate of star formation in the local universe to be maintained, galaxies must refuel the gas reservoirs. How exactly this refuelling takes place is still an open question, however, it has been shown that minor mergers alone cannot supply the required amount of gas necessary to sustain the observed levels of star formation \citep[e.g.][]{DiTeodoro2014}, meaning that some form of gas accretion must be ongoing \citep{Larson1972,Sancisi2008, Ho2019}. There are a number of different ways this accretion of gas can occur: gas-rich mergers, accretion of recycled gas \citep[``fountain model'' e.g.][]{Fraternali2002,Melioli2008,Melioli2009}, or accretion of gas from the cosmic web \citep[][]{VandeVoort2011,Wetzel2015,Cornuault2018,Ho2019,Iza2022}. 

Simulations have shown that accretion of fresh gas into galaxies generally comes in two modes \citep{Keres2004,Keres2009,Nelson2013,Nelson2015,Huang2019}. In ``hot mode'' accretion, the gas is shock heated as it collides with the hydrostatic hot halo near the virial radius, if the gas reaches high enough densities, it can cool and settle onto the disk. However, in ``cold mode'' accretion, the gas is accreted along filaments or in clumps that are not shock heated. Accretion along filaments has been shown to be more important when considering how the gas is accreted onto the galaxies \citep[e.g.][]{Keres2009,VandeVoort2011,VandeVoort2012,Hafen2020,Cadiou2022}. {Simulations suggest that these filaments have \hi column densities of $10^{17}$ to $10^{18}\,\cmt$ \citep{VandeVoort2011, VandeVoort2012,VandeVoort2019,Ramesh2023a} which has been impossible to observe with the necessary resolution until very recently. }

Unambiguously identifying accreting gas observationally is extremely difficult due to the low density of the accreting gas and the numerous different possible explanations for ``accretion-like'' features observed in the gas distribution of nearby galaxies. Deep observations of the \hi in nearby spiral galaxies have identified the presence of extra-planar gas (EPG). This provided an opportunity to study the connection been the gas in the halo and the galaxy disk. EPG has now been identified in many spiral galaxies \citep[see review by][]{Sancisi2008, Marasco2019}. In edge-on galaxies, such as NGC 891, the EPG is a clearly visible, extended component lagging in velocity \citep[e.g.][]{Swaters1997,Oosterloo2007}. In studies of more face-on galaxies such as NGC 2403, the presence of the EPG is inferred by the anomalous velocities of the \hi \citep[e.g.][]{Fraternali2001,DeBlok2014,Li2023, Veronese2023}. EPG has been shown to account for $10\text{ to } 20\%$ of the total \hi mass of the host galaxies \citep{Hess2009,Gentile2013, Vargas2017,Marasco2019}. There has been a concerted effort to identify the origin of EPG, particularly in relatively isolated galaxies \citep{Heald2011a, Gentile2013}. While fountain models have been used to explain the presence of EPG in many galaxies \citep[e.g.][]{Li2023}, the fountain models cannot explain all the observed EPG, suggesting other modes of accretion must also be occurring (e.g. NGC 2297, \citealt{Hess2009}). 

In order to understand the relationship between the EPG, disk gas, and star formation, deep, high resolution \hi observations in combination with multiwavelength tracers of past and ongoing star formation are needed for a representative sample of nearby (\numunit{< 20}{\mpc}) galaxies. High resolution (sub kiloparsec scale) \hi observations enable same-scale comparisons between the \hi morphology and kinematics to tracers of star formation activity (e.g. UV and mid- or near-infrared imaging, molecular gas imaging). The high sensitivity is necessary to detect the low column density EPG \hi gas. The limitations of previous generations of radio telescopes meant that surveys set out to tackle these questions had to optimise for either resolution or sensitivity.

Two of these \hi surveys are The \hi Nearby Galaxy Survey \citep[THINGS,][]{Walter2008} and the Westerbork Hydrogen Accretion in LOcal GAlaxieS \citep[HALOGAS,][]{Heald2011a}. THINGS is an \hi survey with the Very Large Array (VLA) of 34 nearby (\numunit{\text{D} < 15}{\mpc}) gas-rich spiral and dwarf galaxies. The survey made use of the VLA in B, C, and D configurations, which provided the high spatial resolution in combination with the high spectral resolution of the VLA correlator, but \hi column density sensitivity was limited to \numunit{\nhi \sim 4.5 \times 10^{20}}{\cmt} at $3\sigma$ over \numunit{20}{\kms} linewidth at the highest resolution of $6''$ \citep{Walter2008}. Some key results from THINGS include detailed analyses by \citet{Bigiel2008} and \citet{Leroy2008} of the correlation between \hi and molecular hydrogen (\htwo) at sub-kpc scales with star formation, showing that star formation efficiency is driven by the presence of molecular gas. \citet{Bigiel2008} also confirmed that there is a surface density at which \hi saturates, and the gas is only molecular. While \citet{Leroy2008} found that where the \hi component is the dominant gas phase, star formation efficiency decreases with increasing radius from the centre of the galaxy. The high resolution imaging also made it possible to detect radial motions related to the EPG component in a smaller sub-sample of THINGS galaxies \citep{Schmidt2016}. Detecting the radial motions of the gas is key to disentangling how gas is transported from the outskirts to the inner star forming regions of galaxies with ongoing accretion.

The first systematic search for evidence of accretion of cold gas onto galaxies was the HALOGAS survey, conducted using the Westerbork Synthesis Radio Telescope, which was designed to detect the low column density \hi and characterise the morphology and kinematics of the detected EPG. The 24 nearby (\numunit{\text{D} < 20}{\mpc}) galaxies chosen were moderately inclined or edge-on as this makes it easier to model and study in detail the morphology and kinematics of the EPG. The deep observations ($10\times 12\,\text{hours}$ per galaxy) provided high sensitivity \hi maps (\numunit{\nhi \sim 1.1\times 10^{19}}{\cmt} at $3\sigma$ for a linewidth of \numunit{20}{\kms} at a resolution of $30''$, \citealt{Heald2011a}). Studies of the HALOGAS galaxies have provided much of what is now known about EPG in galaxies as discussed above. Another one of the important results from HALOGAS has been the identification of clouds of anomalous gas \citep{Heald2014, Kamphuis2022}, similar to the high velocity clouds (HVC) around the Milky Way \citep{Wakker1997}. \citet{Kamphuis2022} also set important limits on the rate at which \hi can be accreted via these clouds showing that it is much lower than the global accretion rate which is estimated to be \numunit{0.2}{\msun/\text{yr}} by \citet{Sancisi2008}.

The limitations of the previous surveys have been overcome by the arrival of MeerKAT \citep{Jonas2016} where it is no longer necessary to choose between sensitivity and resolution. The MeerKAT \hi Observations of Nearby Galactic Objects - Observing Southern Emitters \citep[MHONGOOSE;][]{DeBlok2016a,DeBlok2020} is one of the \mk Large Survey Projects, having been awarded a total of \numunit{1650}{\text{hours}} to observe the \hi in 30 nearby galaxies (\numunit{55}{\text{hours}} per galaxy).
The MHONGOOSE galaxies were selected from an overlap of galaxies observed as part of both the \hi Parkes All Sky Survey \citep[HIPASS;][]{Barnes2001,Meyer2004} and the Survey for Ionization in Neutral Gas Galaxies \citep[SINGG;][]{Meurer2006a}. This gave a first estimate of the total \hi mass of the sources, and ensured the existence of ultraviolet to infrared observations of the galaxies. All 30 galaxies were also required to be within \numunit{30}{\mpc}, and were chosen to avoid dense environments such as the nearby Fornax and Virgo clusters. In this work, we present the new \mk \hi observations of one of the MHONGOOSE galaxies: NGC 5068 (HIPASS J1318-21). 

\begin{figure}[h]
    \centering
    \includegraphics[width=\linewidth]{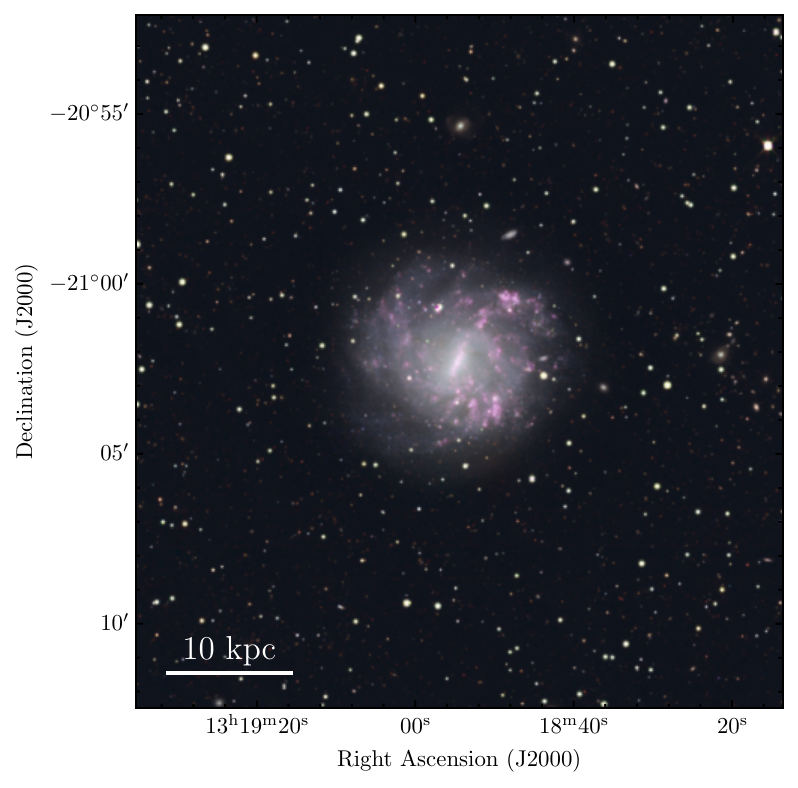}
    \caption{Composite optical/FUV image of \gal. The DECam Legacy Survey \citep[DECaLS;][]{Dey2019} $g$,$r$,$z$ provides the RGB colour. The far UV observed by GALEX as part of the SUNGG Survey \citep{Wong2016} highlights the pink star formation regions.}
    \label{fig:optsf}
\end{figure}

\gal is a nearby, \numunit{d = 5.2}{\mpc} \citep{Anand2021}, close to face-on ($i = 35.7\pm 10.9^\circ$, \citealt{Lang2020a}) barred spiral galaxy \citep{Rosolowsky2021}. It is an actively star forming galaxy of intermediate stellar mass (\numunit{\mstar = 2.29\times 10^9}{\msun}, \citealt{Leroy2019}), located well within the scatter of the so-called star formation ``main sequence''. {\figref{fig:optsf} shows a composite optical/FUV image of \gal which highlights the star-forming regions in the outer regions of the galaxy}. Studies of the \hi in this galaxy, have up until now, only made use of observations by single dish telescopes \citep{Koribalski2004,Sorgho2019c, Sardone2021} which lack the resolution to identify details in the morphology of the \hi disk. This work provides a first look at the high resolution details of the \hi distribution of \gal while still maintaining an \hi column density sensitivity equal to that of the single dish observations.

This paper is laid out as follows: in \secref{sec:data}, we present the new \mk \hi observations of \gal and the process by which we calibrated and imaged the data. In \secref{sec:globalprop} we discuss the global \hi properties of the galaxy, and compare to previously published observations. A detailed discussion on the kinematics of the \hi follows in \secref{sec:kinematics}. Finally, in \secref{sec:discuss}, we explore different possible origins of the anomalous low column density \hi gas detected in \gal.


\section{MeerKAT observations}
    \label{sec:data}

NGC 5068 was observed as part of MHONGOOSE in ten \numunit{5.5}{\text{hour}} tracks (inclusive of time spent on the calibrators) for a total on-source time of \numunit{50}{\text{hours}}. In each track, the primary calibrator, J0408-6545 (rising tracks) or J1939-6342 (setting tracks), was observed for \numunit{10}{\text{min}}. The secondary calibrator, J1311-2216, was observed every \numunit{50}{\text{min}} for \numunit{2}{\text{min}}. {The observations were carried out at night between April 2021 and March 2022 using an average of 62 of the 64 antenna which provide baselines of \numunit{29\text{ to } 7800}{\text{m}}. Each MeerKAT dish has a diameter of \numunit{13.5}{\text{m}} which corresponds to a full width at half max (FWHM) of the primary beam of \numunit{55}{\text{arcmin}} at \numunit{\nu = 1420.405}{\mhz}; for more details see \citet{deBlok2024}.}

Each track was observed using the narrow-band (\numunit{107}{\mhz}) mode of  MeerKAT. With this mode, the observations have a native channel width of \numunit{3.3}{\text{kHz}} or \numunit{0.7}{\kms} at \numunit{1420}{\mhz}. We bin the channels by a factor of two{, to a channel width of \numunit{6.6}{\text{kHz}} or \numunit{1.4}{\kms},} before starting the calibration process. {Due to the shape of the MeerKAT spectral response \citep{VanderByl2022} individual \numunit{0.7}{\kms} channels are independent, implying that our binned \numunit{1.4}{\kms} channels are as well.}\\

\begin{table}[h]
    \renewcommand{\arraystretch}{1.4}
    \centering
    \caption{General properties of \gal.}
    \label{tab:galprop}
    \begin{tabular}{lc}\hline \hline 
        $\alpha$, $\delta$ (J2000) & $13^\mathrm{h}18^\mathrm{m}54.5^\mathrm{s}$ $-21^\circ02\arcmin17\arcsec$ \\
        $v_{sys}$ (\kms) & $667.8 \pm 1.3^a$ \\
        PA ($^\circ$) & $342.4 \pm 3.2^b$ \\
        $i$ ($^\circ$) & $35.7 \pm 10.9^b$ \\
        $D_{25}$ ($\arcmin$) & $7.03^c$ \\
        $d$ (\mpc) & $5.20 \pm 0.22 ^d$ \\
        \mstar (\msun) & $2.29 \times 10^9\,^e$  \\
        SFR (\msun/year) & $0.275 \pm 0.127\,^e$ \\ \hline
    \end{tabular}\\
    \small{Notes: $^a$~calculated from the global profile in this work, $^b$~based on optical properties from \citet{Lang2020a}, $^c$ $B$-band diameter at \numunit{25}{\text{mag
/arcsec}^{2}} isophote from \citet{Lauberts1989}, $^d$~TRGB from \citet{Anand2021}, $^e$~\citet{Leroy2019}}
\end{table}

All 10 tracks were calibrated using the same procedure with version 1.0.5 of the Containerized Automated Radio Astronomy Calibration (\texttt{CARAcal}) pipeline \citep{Jozsa2020}. \texttt{CARAcal} provides a single environment in which to carry out the usual calibration and reduction steps: flagging of data, cross-calibration, splitting of the target, self-calibration, and later the continuum subtraction and spectral line imaging and deconvolution. At each step, the pipeline produces diagnostic plots which help to identify any issues that may arise. These are consolidated outside of \texttt{CARAcal} into a calibration report for each track that is automatically uploaded and stored on a private GitHub team repository where each report is used to quality check the observation.

\subsection{Self calibration and continuum subtraction}
We used self-calibration of the continuum  to further improve the quality of the calibration. For this we used a continuum image created from the frequency range \numunit{1390\text{-}1422}{\mhz}, {taking care to flag the frequency range covered by the \hi emission from both the Galaxy and the target, \gal}. Four rounds of self-calibration were used, where in each round a progressively lower signal-to-noise (S/N) threshold is employed to create a mask using the Source Finding Application \citep[SoFiA,][]{Serra2015}, which thus increases the number of sources to be included in the progressively better sky model. The continuum emission of \gal is bright and extended, and we note the importance of including all of the diffuse continuum emission of \gal in the sky model. Not doing so leads to the appearance of ``ghosts'' in the data, mimicking faint \hi structures \citep{Grobler2014,Wijnholds2016}. The model is also used in the first continuum subtraction step for the \hi spectral line cube. Any residual continuum emission is subtracted using a 1st order polynomial fit to the line-free channels.

\subsection{\hi imaging}

The \hi cubes were imaged using \texttt{WSClean} \citep{Offringa2014,Offringa2017} as part of \texttt{CARAcal}. We employed a 3-step iterative strategy to cleaning the \hi data: first, a low-resolution cube was created and cleaned using a 5$\sigma_{rms}$ clip \texttt{WSClean} auto-mask. From this low-resolution cube, a mask of the source was created using \sofia \citep{Westmeier2021}; second, using the newly created low resolution mask for the cleaning, we created a new cube at the desired resolution from which an updated clean-mask was created using \sofia; in the third step, the final \hi cubes that are presented in this work were created and cleaned using the updated mask from the previous step. In each step, we cleaned the data to $0.5\sigma_{rms}$. This procedure is described in more detail in the MHONGOOSE paper presenting the full survey sample \citep{deBlok2024}.

We make full use of MeerKAT's resolution and sensitivity capabilities by creating a set of six \hi cubes with a $1.5^\circ$ field of view (FoV) that range in resolution ($7\arcsec$ to $90\arcsec$ which correspond to \numunit{0.34}{\kpc} to \numunit{4}{\kpc}). This is done using different combinations of robust weighting ($r$) and Gaussian tapering ($t$), which varies the noise and resolution, and thus the \hi column density sensitivity (\numunit{\nhi \sim 6.8 \times 10^{19}}{\cmt} to \numunit{6.4 \times 10^{17}}{\cmt} at $3\sigma$ over \numunit{20}{\kms} -- see \tabref{tab:sens}). The combinations are listed in the left-most column of \tabref{tab:sens}. Using this broad range in sensitivity and resolution, we can characterise the morphology of the \hi. Also listed for completeness in \tabref{tab:sens} is the resolution (second column), the $\sigma_\mathrm{rms}$ based on \numunit{1.4}{\kms} wide channels (fourth column), $3\sigma$ over \numunit{20}{\kms} column density sensitivity (fifth column), and the column density at S/N$=3$ in the total intensity maps (sixth column) for each cube.

\begin{table*}
    \renewcommand{\arraystretch}{1.4}
    \centering
    \caption{Properties of the \hi cubes.}
    \label{tab:sens}
    \begin{tabular}{lccccc}\hline \hline
        Cube & Resolution & Pixel size & $\sigma_\mathrm{rms}^a$ & \nhi ($3\sigma$ over $20\,\kms$)$^c$ & \nhi (S/N$ = 3$)$^d$\\ 
         & ({arcsec} $\times$ {arcsec}) & (arcsec) &  (mJy/beam) & (\cmt) & (\cmt) \\ \hline
        $r=0.0$, $t=0''^b$ & $8.1 \times 6.9$ & 2 & 0.215 & $6.8 \times 10^{19}$ &      $9.1 \times 10^{19}$ \\ 
        $r=0.5$, $t=0''$ & $13.4 \times 9.4$ &  3 & 0.169  &    $2.3 \times 10^{19}$ &  $3.1 \times 10^{19}$ \\
        $r=1.0$, $t=0''$ & $26.0 \times 17.9$ & 5 &  0.148  & $5.6 \times 10^{18}$ &    $6.5 \times 10^{18}$ \\
        $r=1.5$, $t=0''$ & $34.3 \times 25.6$  & 7 & 0.153 &    $3.1 \times 10^{18}$ & $3.4 \times 10^{18}$ \\
        $r=0.5$, $t=60''$ & $65.3 \times 63.8$  & 20 & 0.241 & $1.0 \times 10^{18}$ & $1.1 \times 10^{18}$ \\
        $r=1.0$, $t=90''$ & $93.8 \times 91.7$  & 30 & 0.313 & $6.4 \times 10^{17}$ & $7.0 \times 10^{17}$ \\ \hline
    \end{tabular}\\
    \small{Notes: $^a$ this is measured per \numunit{1.4}{\kms} channel. $^b$ $r$ is the robust weighting, and $t$ the Gaussian taper. $^c$ based on $3\sigma$ detection in 14 channels. $^d$ mean column density at the S/N=3 contour in the moment 0 map, see Section 3.1 for details.}
\end{table*}

 \section{Global \hi properties}
\label{sec:globalprop}

\subsection{Morphology of the \hi disk}

    \begin{figure}[h]
        \centering
        \includegraphics[width=0.99\linewidth]{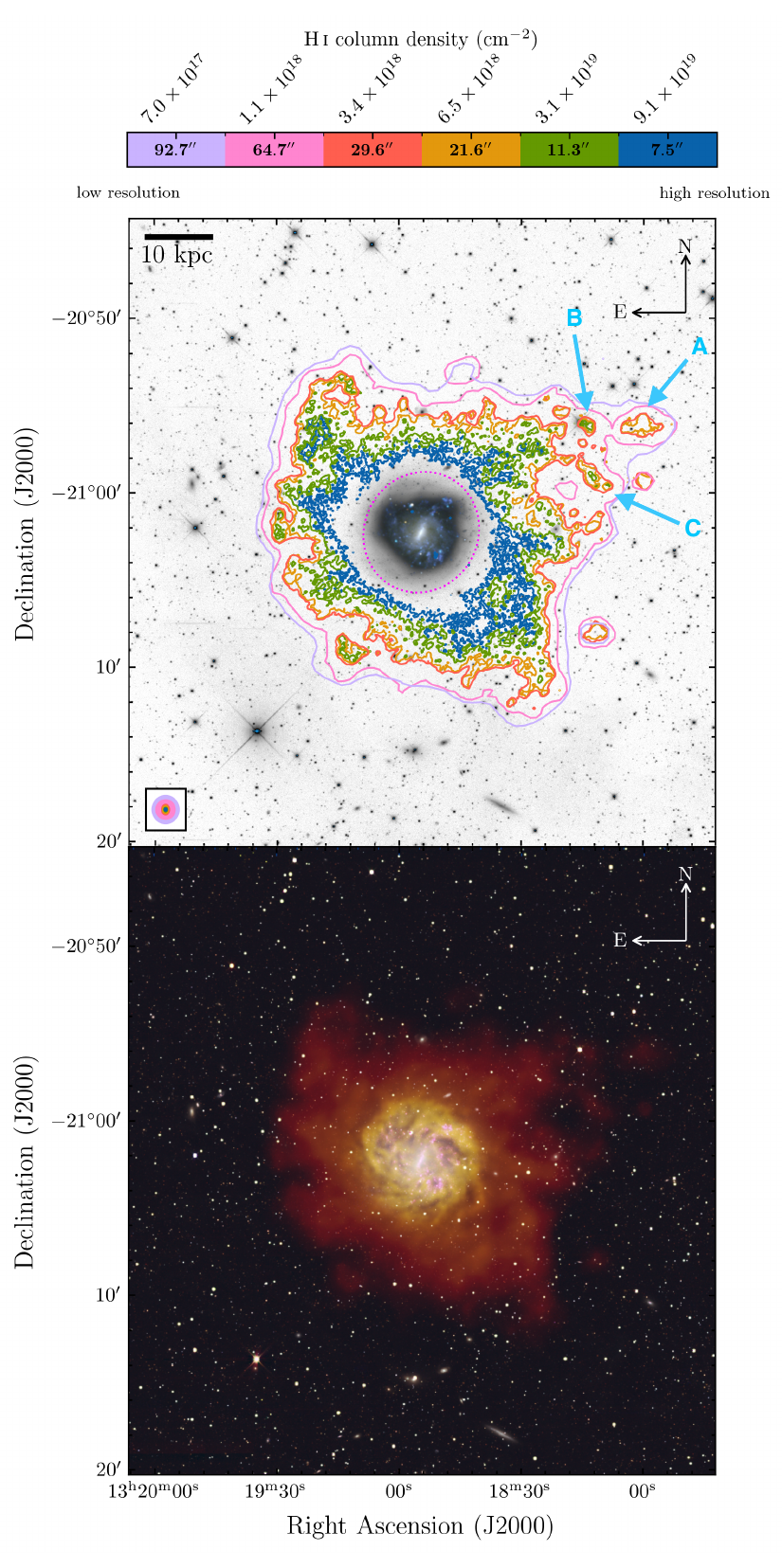}
        \caption{\textit{Top:} Greyscale MeerLICHT $q$-band image and colour with contours from the six different resolution \hi cubes. The contours correspond to colours in the colourbar which indicate the resolution and the \hi column density at the S/N=3 contour in the intensity maps. Three clouds associated with the galaxy are labelled A,B,C. Dotted magenta ellipse indicates the optical extent of the galaxy. The beam for resolution (contour) is shown in the corresponding colour in the box in the bottom left corner. {\textit{Bottom:} \hi emission combined with \figref{fig:optsf}. The red, orange, and yellow show moment 0 maps of the \hi emission from the $64.7''$ (\texttt{r05\_t60}), $29.6''$ (\texttt{r10\_t00}), and $7.5''$ (\texttt{r00\_t00}) resolution \hi cubes respectively. Production of this image followed \citet{English2017}, including the technique of masking in order to combine the data sets and reveal the optical data’s H{$\,\textsc{ii}$} regions (light pink).}}
        \label{fig:mom0_nhi}
        \vspace{-14pt}
    \end{figure}

    \begin{figure}
        \centering
        \includegraphics[width=\linewidth]{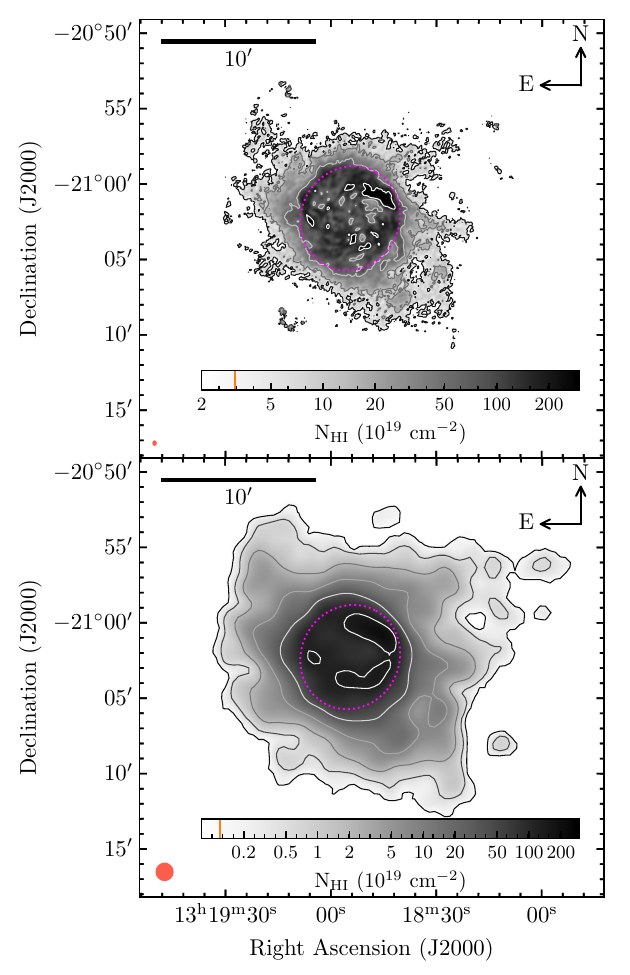}
        \caption{Integrated intensity (moment 0) maps of \gal. In the top panel is the map from one of the high resolution cubes (\texttt{r05\_t0}) at $11\arcsec$, and in the bottom panel is a map from one of the low resolution cubes (\texttt{r05\_t60}) at $64\arcsec$. The dashed magenta ellipse is centred on the optical centre of the galaxy and represents the extent of the optical disk. The contours increase as $\sigma_{S/N=3} \times 2^n,\,\, n=0,2,4,6,...$. The lowest contour plotted is the S/N$=3$ contour, \numunit{\nhi = 3.1 \times 10^{19}}{\cmt} for the top panel and \numunit{\nhi = 1.1 \times 10^{18}}{\cmt} for the bottom panel, {these S/N$=3$ ($n=0$)} values are indicated on the respective colour bars by the orange line. The red ellipse in the bottom left corner of each panel represents the beam. }
        \label{fig:mom0maps}
    \end{figure}

    \begin{figure*}[!h]
        \centering
        \vspace{-10pt}
        \includegraphics[width=\linewidth]{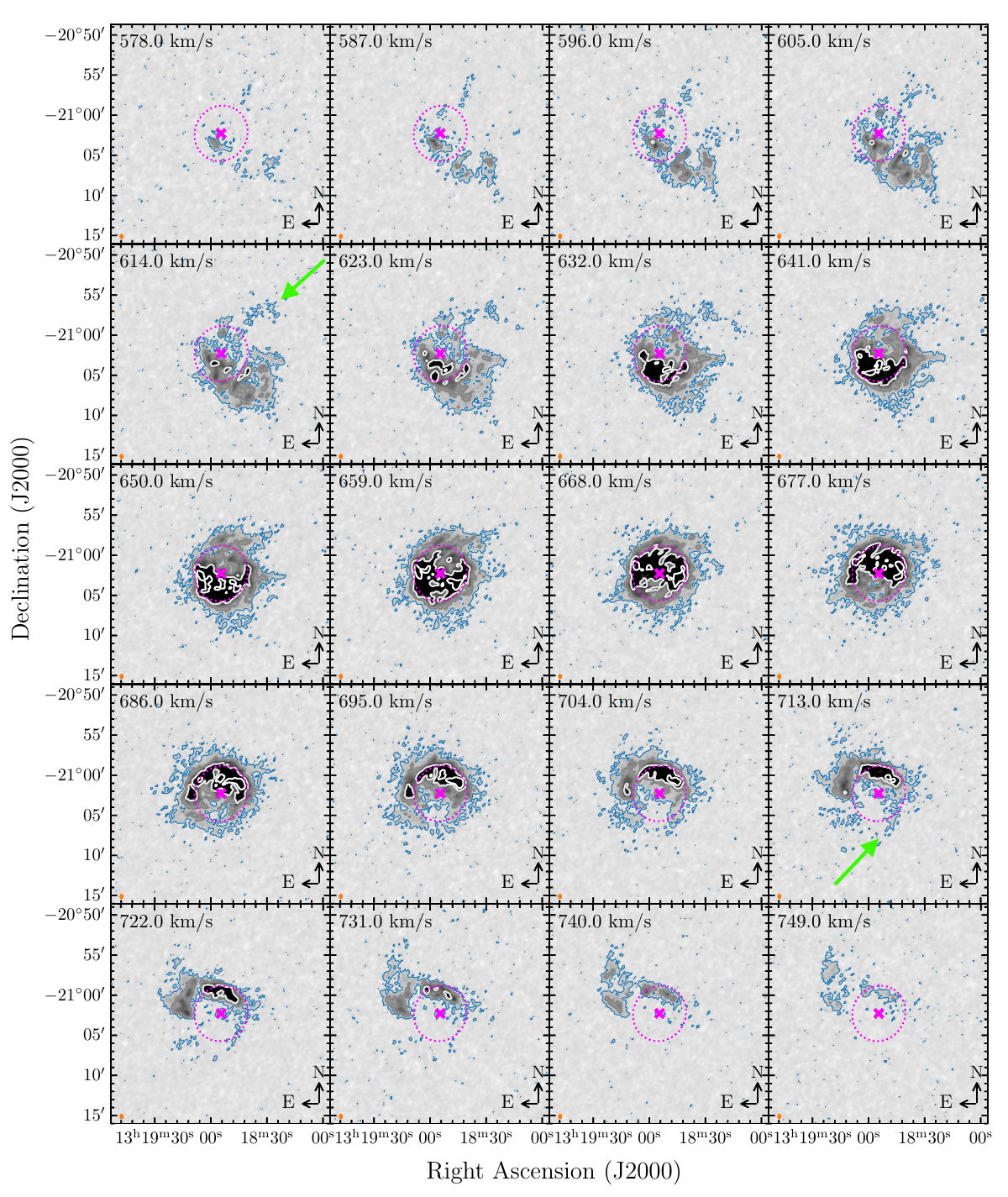}
        \vspace{-15pt}
        \caption{Channel maps taken from the \texttt{r10\_t0} cube that has been smoothed with a 5 channel Hanning window and regridded to \numunit{3}{\kms} channels. Every third \numunit{3}{\kms} channel for a \numunit{170}{\kms} range centred on the systemic velocity is shown. The blue contour indicates 3$\sigma_{rms}$, the grayscale contours are $3\sigma_{rms} \times 2^n,\,\,n=2,4,6,..$. The magenta cross indicates the optical centre of the galaxy, and dotted magenta ellipse indicates the extent of the optical disk ($D_{25}$) of the galaxy. The green arrows point to ``fingers'' in the \hi. The orange ellipse in the lower left corner of each panel represents the $21.6\arcsec$ beam.}
        \label{fig:channelmaps}
        \vspace{-12pt}
    \end{figure*}

    The six \hi cubes enable us to explore both the details of the dense \hi in the galaxy disk, as well as the low column density EPG. For each cube we create a suite of moment maps by using \sofia to identify all detectable \hi emission in the velocity range ($200 < v_{rad}\,(\kms) < 1143$) covered by the data cube. We used the smooth and clip algorithm to identify all signal above a threshold of $3.5\sigma$ using using all combinations of the following spatial and spectral smoothing kernels: spatially, these are the original resolution and a spatial smoothing kernel equal to the synthesised beam; spectrally, we used the original resolution and kernels of 9 channels (\numunit{\sim 12}{\kms}), and 25 channels (\numunit{\sim 35}{\kms}). The source finding yields the \hi emission associated with only the target galaxy, \gal, within a field of view of $1.5^\circ$ which corresponds to a radius of \numunit{125}{\kpc}. The primary beam\footnote{The MeerKAT primary beam FWHM is $55\arcmin$ at $z=0$, this entirely contains the \hi of \gal which spans roughly $30\arcmin$.} corrected moment maps are created from the masked data cubes, however we further limit the first moment (intensity-weighted mean velocity field) and second moment (sometimes referred to as the velocity dispersion map) maps to only include the pixels where the integrated \hi column density (\nhi) is greater than the $\mathrm{S/N}=3$ threshold. This threshold is determined for each resolution map by using the S/N map output by \sofia where $\mathrm{S/N} = \sum S\, \mathrm{d}v/(\sigma_{rms} \mathrm{d}v\, \sqrt{N_{chan}})$, $S\, \mathrm{d}v$ is the moment zero, $\sigma_{rms}$ is the noise per channel of the \hi cube, $\mathrm{d}v$ is the channel width, and $N_{chan}$ is the number of channels of the \hi cube that contributes to each pixel of the zeroth moment map. This is only valid under the assumption that the channels are independent which in this case is true as the observed channels were binned in pairs and no Hanning smoothing was used. Following \citet[][Eq. 75]{Meyer2017}, we calculate the $\mathrm{N_{\hi,\, S/N=3}}$ threshold by calculating the median column density of the pixels which are in the range $2.75 < \mathrm{S/N} < 3.25$ in the S/N map. Despite the differences in how the values are calculated, the $\mathrm{N_{\hi,\, 3\sigma}}$ and $\mathrm{N_{\hi,\, S/N=3}}$ values presented in \tabref{tab:hiparam} column 5 and 6 respectively, are similar although the $\mathrm{N_{\hi,\, S/N=3}}$ values tend to be slightly higher. This is due to a combination of the smooth and clip algorithm used to identify emission associated with a detection, and the linking lengths used in SoFiA-2 to add pixels containing emission to the detection mask which results in a mask that is wider than what would be created using a simple $3\sigma$ clip. It is worth noting that the column density limits presented in \tabref{tab:hiparam} are significantly lower than the previous generation of \hi column density limits, and reflect the excellent quality of these deep MeerKAT data.

    The top panel of \figref{fig:mom0_nhi} shows the \nhi(S/N$=3$) contours of the \hi distribution seen in the different resolutions plotted on the MeerLICHT \citep{Bloemen2016} $q$-band\footnote{The MeerLICHT $q$-band is a wide-band that covers the $g$ and $r$ bands.} image which has a $5\sigma$ surface brightness limit of \numunit{24.61}{\text{mag}/\text{arcsec}^{2}}. MeerLICHT is a 0.6m optical telescope, located in Sutherland, South Africa, with a $1.6^\circ\times1.6^\circ$ FoV ({$\sim1.7\times$ FWHM of the MeerKAT primary beam} at $z=0$). The bottom panel of \figref{fig:mom0_nhi} combines \hi moment maps with \figref{fig:optsf} and shows spatial relationships between the star-forming regions and atomic gas structures in the form of \hi densities or cavities. The upper panel shows the extent of the gas at different column density levels, with the low column density gas extending out to more than 2.5 times the optical radius, defined by the $B$-band \numunit{25}{\text{mag}/\text{arcsec}^{2}} isophote \citep{Lauberts1989}, of the galaxy which is approximately \numunit{10}{\kpc}.  \figref{fig:mom0_nhi} also shows finger-like tendrils of \hi extending out from the optical disk at column densities of \numunit{\nhi \sim 3.1\,\text{to}\,9.7 \times 10^{19}}{\cmt} (green contour). Also of interest are the low-column density clumps on the north-western side of the galaxy which were identified by eye. These clumps appear as distinct clouds below \hi column densities of \numunit{\nhi \sim 3 \times 10^{19}}{\cmt} but become connected to the galaxy \hi disk in the $60\arcsec$ (pink) and $90\arcsec$ (purple) resolution cubes which is likely a resolution effect. The more prominent clumps have been labelled A, B, C in the top panel of \figref{fig:mom0_nhi}. While \figref{fig:mom0_nhi} provides an overall picture of the total \hi intensity distribution, to be able to study the different features, it is necessary to examine the individual moment 0 maps at the various resolutions.

    Figure~\ref{fig:mom0maps} shows the moment 0 (\hi intensity) maps at two different resolutions: $11\arcsec$ (\numunit{0.5}{\kpc}) and $64\arcsec$ (\numunit{3}{\kpc}). In the top panel the high resolution (\texttt{r05\_t0} which corresponds to the robust $(r)=0.5$ and taper $(t)=0\arcsec$ cube) map highlights the clumpy nature of the \hi, in particular the finger-like tendrils and the clouds on the outskirts of the galaxy. The optical disk is represented by the dotted magenta ellipse measured as the diameter of the \numunit{25}{\text{mag}/\text{arcsec}^{2}} isophote ($D_{25}$) in the $B$-band by \citet{Lauberts1989}. In the bottom panel of \figref{fig:mom0maps}, the lower resolution (\texttt{r05\_t60}) map shows the extent of the low column density gas which connects the clouds visible in the higher resolution map to the \hi in the galaxy.

    The individual channel maps provide a clearer picture of the fingers and clouds. \figref{fig:channelmaps} shows a selection of channels taken from the $r = 1$, no taper (\texttt{r10\_t0}) cube, which has a resolution of $21.6\arcsec$, that has been Hanning-smoothed with a 5 channel kernel and regridded to a channel width of \numunit{3}{\kms} to increase the signal to noise of the lowest column density features. This smoothed version of the \texttt{r10\_t0} cube provides a good balance between resolution and column density sensitivity. We note two distinct features that are evident in the channel maps: the first being two ``fingers'', one which points out towards the north west (NW) and the other pointing towards the south east (SE) -- both indicated by the green arrows in the Figure. The second feature is that there appears to be two disks: one inner disk that has a major axis running approximately NW-SE and is coincidental with the optical disk, and the second outer disk which appears to be more elliptical and has a major axis running approximately NE-SW.

    The global profile, extracted from the $r=1.0$, $t=90\arcsec$ (\texttt{r10\_t90}) cube using the associated \sofia mask, is shown in panel (a) of \figref{fig:globalprofile}. For each resolution cube, we determine the mass encapsulated by the mask, the values are shown in panel (b). It is clear that the \hi mass does not change significantly between the largest four resolutions which suggests that all the \hi mass for the system has been identified. The robust weighting used to create the lower resolution \hi cubes emphasises the shorter baselines making these measurements more suitable for computing the total \hi mass. 
    
    From the global profile in \figref{fig:globalprofile}, we calculated the total integrated flux, the line width at 50\% ($w_{50}$) and 20\% ($w_{20}$) of the peak flux are measured directly from the profile, and we determine the systemic velocity to be the velocity at the mid point of the $w_{20}$. These values are listed in \tabref{tab:hiparam}. While the largest contribution to the flux uncertainty are from the systematics arising as part of the calibration process, the uncertainties on the global profile are calculated base on the noise properties of the corresponding \hi cube.

 \begin{table}
    \renewcommand{\arraystretch}{1.4}
    \caption{Global \hi properties of \gal measured from the global profile.}
    \label{tab:hiparam}
    \centering
    \begin{tabular}{lc}\hline \hline
       Parameter  & Value \\ \hline
       $S_{int}$ & \numunit{167.82 \pm 0.15}{\text{Jy}\,\kms} \\
        \mhi (total) & \numunit{1.07 \pm 0.09 \times 10^9}{\msun} \\
        $w_{50}$ & \numunit{67.1 \pm 0.6}{\kms} \\
        $w_{20}$ & \numunit{109.1 \pm 2.0}{\kms} \\
        $v_{sys}$ & \numunit{667.8 \pm 1.3}{\kms}\\[5pt]
        \hline
    \end{tabular}
\end{table}

\subsection{Comparison with previous observations}

\begin{figure}
    \centering
    \includegraphics[width=\linewidth]{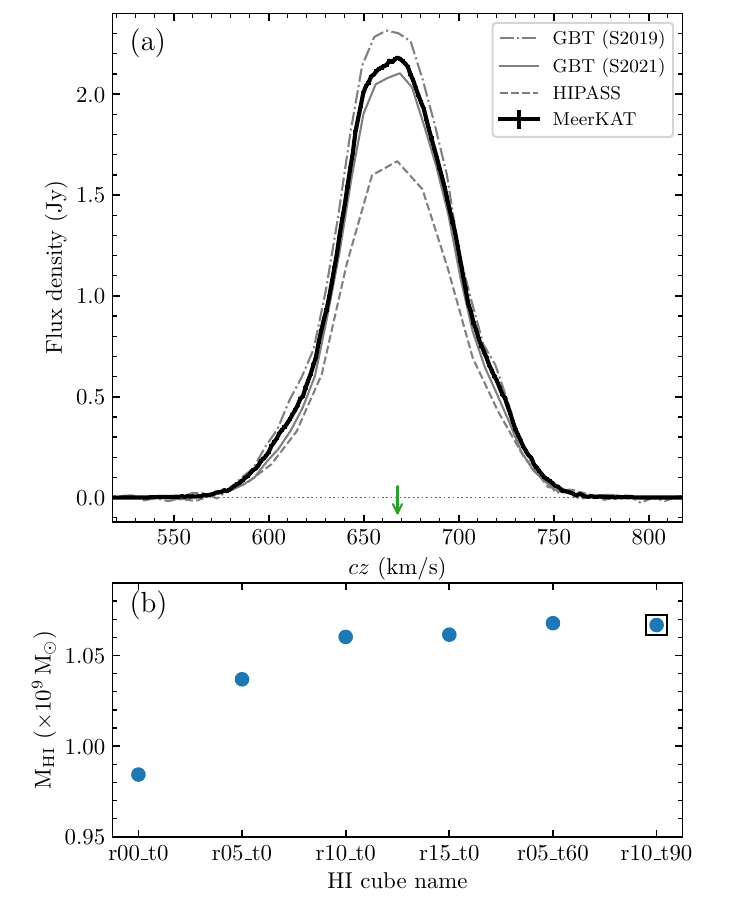}
    \caption{(a) Global \hi profile of \gal from MeerKAT (black), HIPASS (dashed grey; \citealt{Koribalski2004}), GBT (solid grey and dot-dash grey; \citealt{Sorgho2019c, Sardone2021}). The green arrow indicates the systemic velocity of the galaxy. (b) The \hi mass of \gal measured within the \sofia mask for each different resolution cube. The open black square indicates which cube was used to create the global profile shown in the top panel.}
    \label{fig:globalprofile}
\end{figure}

    Here we compare the total \hi flux of \gal with those obtained from single-dish measurements from the \hi Parkes All Sky Survey \citep[HIPASS,][]{Meyer2004,Koribalski2004} and targeted observations using the Green Bank Telescope \citep[GBT][]{Sorgho2019c,Sardone2021}. \citet{Sorgho2019c} and \citet{Sardone2021} analysed a sub-sample of the MHONGOOSE galaxies using GBT. Although they use the same GBT spectra, they process the data differently which resulted in slightly different global profiles and measurements of the integrated flux for \gal. The total integrated flux measurements from each survey are listed in \tabref{tab:intflux}, and the global profiles are also plotted in \figref{fig:globalprofile}(a). 
 
    As is evident from \tabref{tab:intflux}, there are differences in measured fluxes between the different single-dish measurements. Despite the differences in the integrated fluxes, the \hi global profile line widths ($w_{20}$ and $w_{50}$) measured for \gal by \citet{Sorgho2019c} and \citet{Sardone2021} from the GBT observations, and by \citet{Koribalski2004} from HIPASS are consistent with what we measure from the \mk spectra. 
    
    While {the shape and amplitude of the} \citet{Sardone2021} global profile is consistent with the \mk spectrum, the quoted integrated flux is {inconsistent}. {The higher flux density measurement quoted in \citet{Sardone2021}} {is due to how the value was calculated: they determined the total flux density of the galaxy by measuring the flux in annuli with increasing radius from the galaxy centre, and then summing the flux in all the annuli. This method of determining the total flux of the galaxy is not the same as the} integrated flux based on the global profile. We recalculated the integrated flux from the \citeauthor{Sardone2021} global profile to be \numunit{S_\hi = 157.7}{\text{Jy}\,\kms} which is consistent within 10\% of the \mk value. 
    
    Given that we do not expect the HIPASS observations of \gal to missing any flux, it is therefore curious that there is such a large discrepancy between the total integrated flux measured from the \mk spectrum and that which is reported by \citet{Koribalski2004}. However, the HIPASS data are known to underestimate the \hi emission of large galaxies as a result of the bandpass correction method \citep{Barnes2001} which likely subtracted some of the \hi (see \citet{deBlok2024} for a detailed comparison with all 30 MHONGOOSE galaxies).

    \begin{table}
        \renewcommand{\arraystretch}{1.4}
        \centering
        \caption{The total integrated flux and linewidth measurements for \gal from different \hi surveys}
        \label{tab:intflux}
        \resizebox{\hsize}{!}{
        \begin{tabular}{lccc}\hline \hline
            Survey & Integrated flux  & $w_{20}$ & $w_{50}$  \\ 
                    &  (Jy$\,$\kms) &  (\kms) &  (\kms) \\ \hline
            HIPASS \citep{Koribalski2004} & $128.6 \pm 12.7$ & 111.0 & 70.0\\
            GBT \citep{Sorgho2019c} & $177.0 \pm 0.7$ & -- & 69.3 \\
            GBT \citep{Sardone2021} & 191.5$^a$  & 108.0 & 67.9 \\ 
            MeerKAT (this work) & $167.82 \pm 0.15$ & $109.1 \pm 2.0$ & $67.1 \pm 0.6$ \\[5pt] \hline
        \end{tabular}}\\
    {\small{Notes: $^a$ \numunit{S_\hi = 157.7}{\text{Jy}\,\kms} measured from the global profile assuming a gain of \numunit{1.86}{\text{K/Jy}} from \citet{Sardone2021}} }
    \end{table}

\section{Kinematics of the \hi}
    \label{sec:kinematics}
    \begin{figure}
        \centering
        \includegraphics[width=\linewidth]{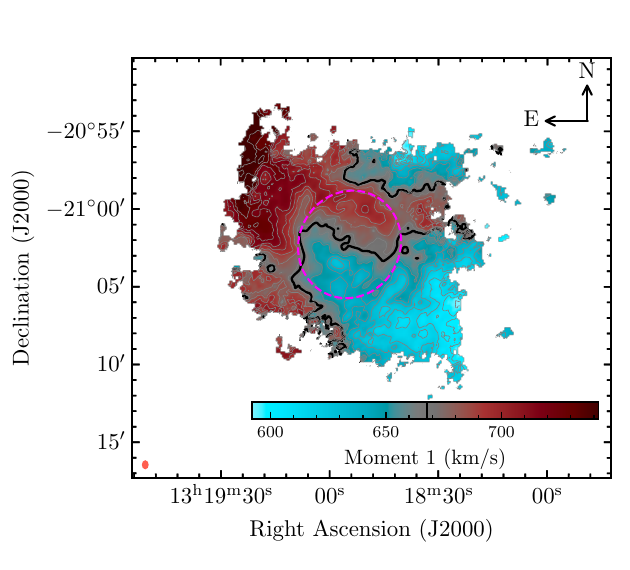}
        \caption{Velocity field (moment 1) for \gal. Pixels with an \hi column density below \numunit{\nhi = 5.9 \times 10^{18}}{\cmt} (S/N=3) are masked. The colourbar in the bottom of the image indicates the velocity of the gas, the grey contours are spaced \numunit{10}{\kms} apart with the black contour indicating the systemic velocity, \numunit{v_{sys} = 667}{\kms}. The dotted magenta ellipse is centred on the optical centre of the galaxy and represents the optical size of the galaxy.}
        \label{fig:moment1}
    \end{figure}

    \begin{figure}
        \centering
        \includegraphics[width=\linewidth]{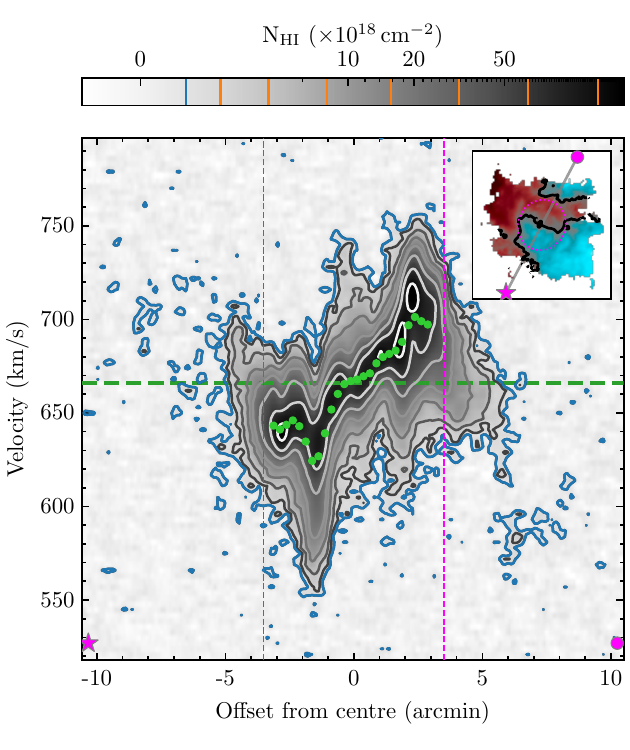}
        \caption{Position-velocity diagram extracted with a width of $21.6\arcsec$ which corresponds to the size of the beam. The slice is extracted along optical major axis. The inset on the top right of the figure shows the velocity field from \figref{fig:moment1}, the grey line running from the magenta star to the filled circle represent the path along which the PV slice was extracted, the magenta star and circle are located in the lower left and upper right corners indicating the direction of the slice, these symbols are repeated lower left and right corners of the main figure. The horizontal dashed green line indicates the systemic velocity, while the light green filled circles show the moment 1 velocity at each position along the major axis. The dashed vertical magenta lines correspond to the edge of the optical disk represented by the dotted magenta ellipse in the inset. The blue and orange contours in the colour bar correspond to the blue and greyscale contours on the PV slice.}
        \label{fig:pvmajor}
    \end{figure}

\begin{figure}
        \centering
        \includegraphics[width=\linewidth]{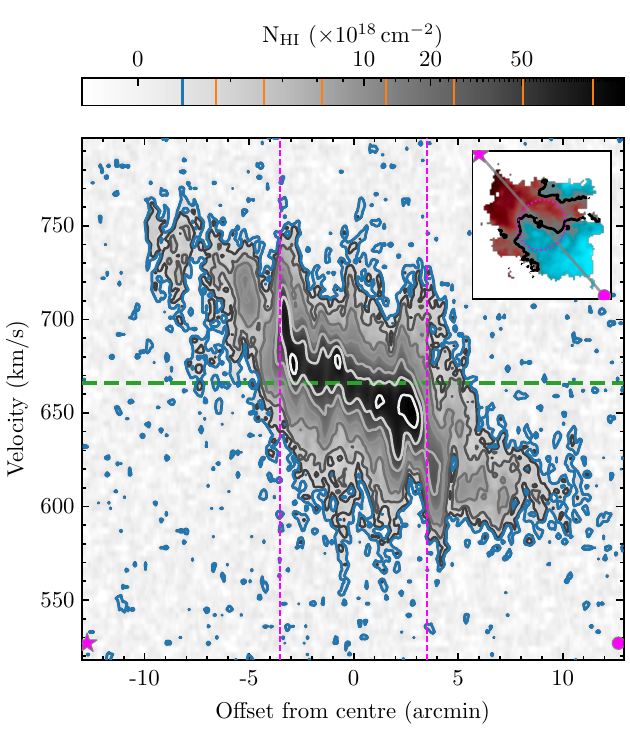}
        \caption{Same as \figref{fig:pvmajor} but at an angle of $224^\circ$ through the outer disk.}
        \label{fig:warppv}
    \end{figure}

    \begin{figure}
        \centering
        \includegraphics[width=\linewidth]{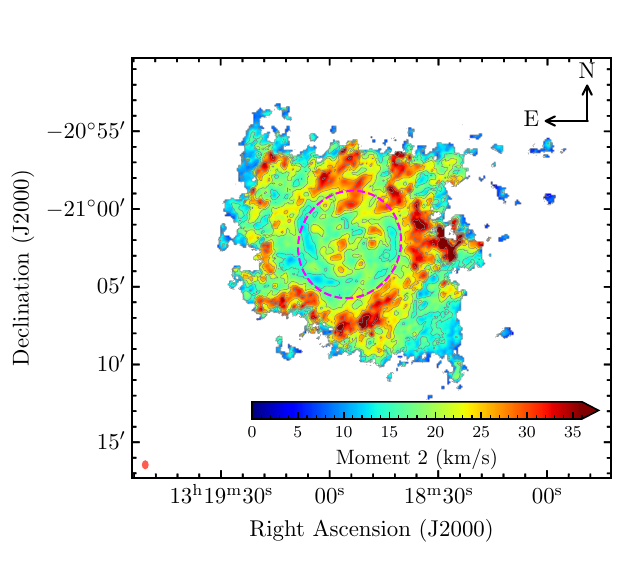}
        \caption{Moment 2 map for \gal at $21.6\arcsec$ resolution. Pixels with an \hi column density below the S/N=3 threshold of \numunit{\nhi = 5.9 \times 10^{18}}{\cmt} are masked. The colourbar in the bottom of the image indicates the moment 2 velocity of the gas, the thin grey contours are spaced \numunit{5}{\kms} apart. The dotted magenta ellipse is centred on the optical centre of the galaxy and represents the optical size of the galaxy.}
        \label{fig:moment2}
    \end{figure}

    \begin{figure*}
        \centering
        \includegraphics[width=\linewidth]{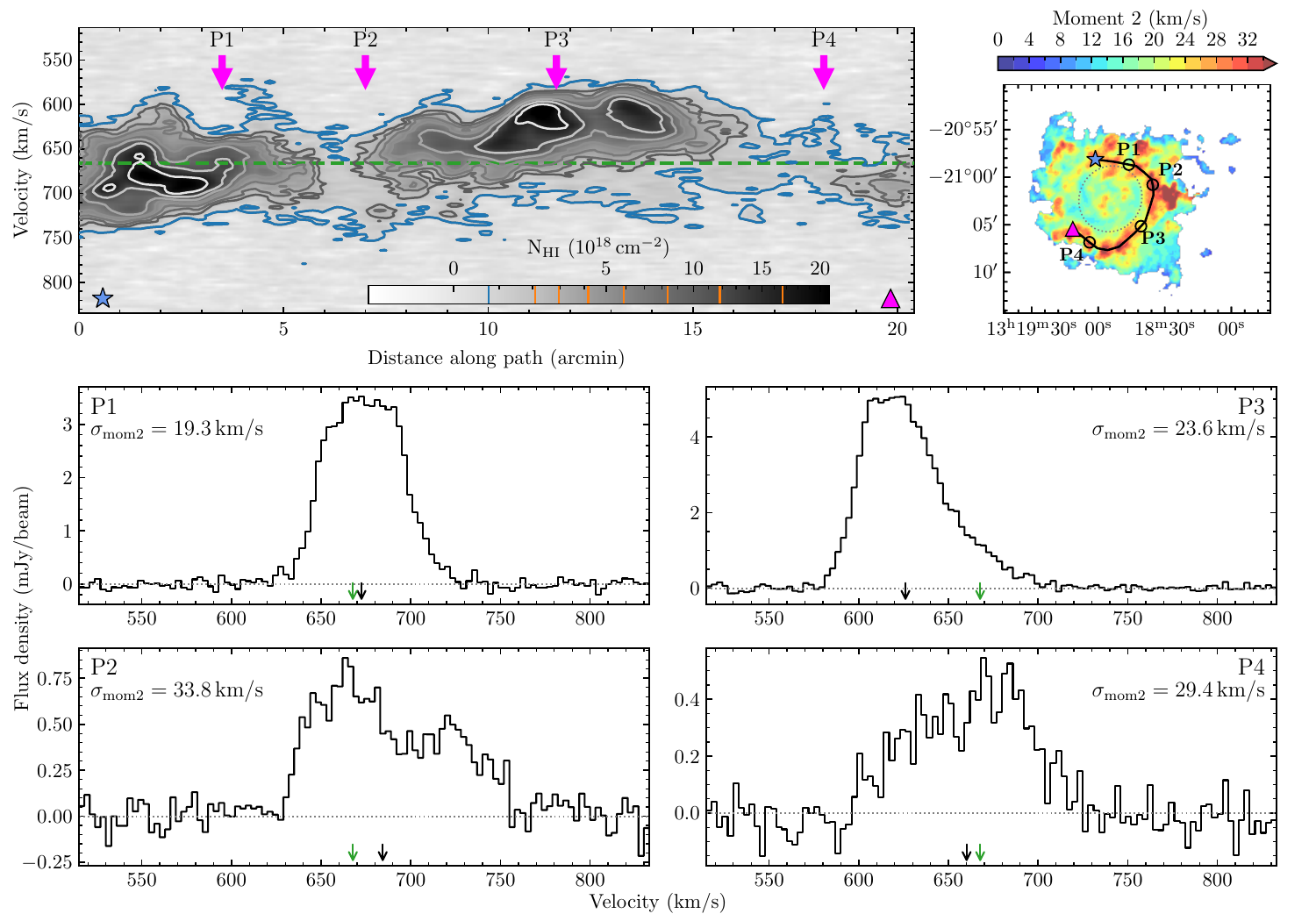}
        \caption{Top left: Position-velocity (PV) diagram through the high moment 2 region of \gal. The blue and orange lines in the grayscale colourbar correspond to the blue and grayscale contours. The green dashed line indicates the systemic velocity. Top right: moment 2 map of the galaxy with the black line tracing the path along which the PV slice was extracted, starting from the blue star and ending at the magenta triangle. Bottom: \hi line profiles extracted at different locations along the PV slice, the extracted locations are indicated by the magenta arrows in the PV diagram. The location of each spectrum is also indicated by the labelled open circles on the moment 2 map. Green arrows at the bottom of the line profiles indicate the systemic velocity of the galaxy, while the black arrows point to the moment 1 velocity. }
        \label{fig:pvslice}
    \end{figure*}

    In \figref{fig:moment1}, we present the intensity weighted mean velocity field (moment 1) from the \texttt{r10\_t0} cube at $21.6\arcsec$. For this analysis, we use the \texttt{r10\_t0} cube as it provides a good combination of resolution and \hi column density sensitivity. The systemic velocity ($v_{sys}$, listed in \tabref{tab:hiparam}) is measured as the central velocity of the global profile -- this is a reasonable assumption given the symmetry of the global profile. The $v_{sys}$ is indicated by the green arrow in \figref{fig:globalprofile}, and by the thick black contour in \figref{fig:moment1}. The \hi kinematics presented in  \figref{fig:moment1} are rather complex, but there appears to be at least three main components to the velocity field: (1) a regularly rotating inner disk with a mass measured within the \numunit{\Sigma_\hi = 1}{\msun\,\text{pc}^{-2}} contour of \numunit{\mathrm{M_{disk}} = 9.7 \times 10^8}{\msun}; (2) a separate, inclined warped ``disk'' which extends to larger radii than the optical disk, that has an \hi mass of \numunit{\mathrm{M_{outer} = 8.9 \times 10^7}}{\msun}, calculated as all the \hi outside the inner disk (including the 3rd component); (3) the north/north western quadrant which is home to the clumpy clouds identified  in \figref{fig:mom0_nhi} which make up on average \numunit{8.9 \times 10^4}{\msun} each, with the most massive being C at \numunit{1.6 \times 10^5}{\msun}.

    \begin{table}
        \renewcommand{\arraystretch}{1.4}
        \centering
        \caption{\hi mass measurements of the different components of the \hi disk of \gal.}
        \label{tab:hicomp}
        \begin{tabular}{lc} \hline
            Component & \hi mass ($\msun$) \\ \hline
            M$_\mathrm{\hi,\, total}$ & $1.07 \times 10^9$ \\
            M$_\mathrm{disk}$ & $9.7 \times 10^8$ \\ 
            M$_\mathrm{outer}$ & $8.9 \times 10^7$\\
            Average M$_\mathrm{clumps}$ & $8.9 \times 10^4$ \\
            \mhi (residual clumpy gas) & $2.6 \times 10^7$ \\
            \hline
        \end{tabular}
    \end{table}

    The first component is the regularly rotating main inner disk that is coincident with the optical body represented by the magenta ellipse in \figref{fig:moment1}. We find that the \numunit{\Sigma_\hi = 1}{\msun\,\text{pc}^{-2}} contour {(which is roughly equivalent to \numunit{1.2 \times 10^{20}}{\cmt}, the second lowest contour in the top panel of \figref{fig:mom0maps})} neatly encircles this region of the velocity field. This contour {(\numunit{\Sigma_\hi = 1}{\msun\,\text{pc}^{-2}})} is usually used to determine the \hi diameter of galaxies \citep[e.g.][]{Wang2016}, a parameter that has a very tight correlation with the \hi mass. The contour is also roughly $20\text{ to }50\%$ larger than the optical disk (see {top panel of} \figref{fig:mom0maps}), which is consistent with other measurements of late type galaxies \citep{Bosma2016}. Compared to the total \hi mass of the system (see \tabref{tab:hicomp}), the inner disk (\numunit{M_\mathrm{disk} = 9.7 \times 10^8}{\msun}) clearly makes up the majority ($\sim 90\%$) of the \hi mass of the galaxy.
    
    A position velocity (PV) diagram extracted along the optical major axis (see \tabref{tab:galprop}) is presented in \figref{fig:pvmajor}. The PV slice is extracted using the width of the beam (21.6\arcsec) from the \texttt{r10\_t0} cube that has been Hanning smoothed and regridded to \numunit{3}{\kms}. The PV slice shows that while in the region corresponding to the optical disk (denoted by the vertical magenta lines) there is regular rotation as seen in the velocity field (\figref{fig:moment1}). The light green circles overlaid on the PV slice in \figref{fig:pvmajor} indicate the moment 1 velocity in each line of sight. Anomalous gas is clearly visible in the PV slice, indicated by the regions of gas deviating from the regular rotation and extending to higher and lower velocities. The sudden dip in velocity on the approaching side of the galaxy at roughly $-1.4\arcmin$ offset from the centre has similar characteristics to the \hi holes identified in NGC 6946 \citep{Kamphuis1993,Boomsma2008}, and M31 and M33 \citep{Deul1990}. {While the kinematics of the feature are similar to the NGC 6946 holes, there is no obvious hole in the morphology of the \hi. Thus it is more likely that the feature in the PV diagram is indicative of some kind of expanding feature or shell or a hole that has not blown out yet which is why there is no corresponding feature in the intensity map. Features such as holes or shells} in the \hi distribution are thought to be caused by energetic processes such as supernovae or stellar winds. Since this dip in the PV slice coincides with emission in the continuum and far-UV, both tracers of ongoing and recent star formation, it is likely the aforementioned processes are responsible for the feature. Another noteworthy feature of the PV slice is that there is clearly low column density extraplanar gas distributed throughout the disk. 

    The second component is comprised of the low column density gas (\numunit{\nhi < 10^{19}}{\cmt}) at radii larger than the inner disk. The gas in this component appears to have disk-like kinematics with a position angle of $\sim\!224^\circ$. In \figref{fig:warppv} we present a PV diagram through the centre of \gal at this angle which shows clearly the rotation of this component at larger radii than the inner disk (represented by the vertical magenta lines in the figure). The elliptical geometry of this component suggests that it is more inclined than the inner disk -- approximately $i \sim 53^\circ$. A closer look at the kinematics in \figref{fig:moment1} show that the outer regions are twisted in an S-shape which suggests a warp. Deep $r$- (\numunit{\mu_{3\sigma} = 26.5}{\text{mag}/\text{arcsec}^2}) and $g$-band (\numunit{\mu_{3\sigma} = 27.2}{\text{mag}/\text{arcsec}^2}) imaging from the DECam Legacy Survey \citep[DECaLS,][]{Dey2019} Data Release 10 (DR10) shows that there is no stellar counterpart associated with this gas.

    The differences in the position angle and inclination between this component and the inner disk suggest that this outer disk is not as a result of gas being swept out of the inner disk, but rather it has a separate origin. Given the kinematics and the geometry of the outer disk relative to the inner disk, we describe it as a warped inclined outer disk which could possibly be a warped polar disk. Characteristically polar disks are highly inclined relative to the inner disk and the position angles can also be very different to that of the inner disk -- see the prototypical examples of NGC 4650A \citep[]{Arnaboldi1997} and NGC 660 \citep{Gottesman1990,vanDriel1995}. Without proper modelling (see Deg et al. under review, for detailed modelling of polar disk galaxy candidates), the polar disk scenario is difficult to confirm. Throughout the rest of this work we will primarily refer to this component as the outer disk.

    The northern region of the galaxy containing the clouds marked A,B,C in \figref{fig:mom0_nhi} and \figref{fig:moment1} makes up the third component.    The kinematics of this region are not consistent with either the inner or outer disks, and thus we consider them a separate component. 

    \subsection{High moment 2 ring}
    In \figref{fig:moment2}, we present the second moment map at $21.6\arcsec$ resolution for \gal. The moment 2 can be used as a measure of the linewidth of the \hi along each line of sight. In the case that the line profiles are Gaussian, then the moment 2 represents the velocity dispersion. Normally the velocity dispersion across the star forming disk is on average \numunit{\sim 10 \text{ to } 15}{\kms} \citep[e.g.][]{Leroy2008}. The median moment 2 value within the optical disk region (indicated by the dashed magenta ellipse in \figref{fig:moment2}) is \numunit{\sigma_v \sim 18}{\kms}. Outside the optical disk, there are regions where the moment 2 values are significantly higher (\numunit{\sigma_v > 25}{\kms}) than in the inner disk. There are a number of processes that can add energy to the interstellar medium of galaxies, and thus drive turbulence leading to an increase in the velocity dispersion. Many of these processes are associated with star formation activity, such as stellar feedback and supernova explosions \citep[e.g.][]{Tamburro2009, Krumholz2016}. Since the high moment 2 values in \gal are found outside the stellar disk, it is likely that something other than star formation related processes is responsible for the high moment 2 values.
        
    In order to investigate the source of the high moment 2 values, we take a PV slice 1 beam wide ($21.6\arcsec$) through the ring of high moment 2 values on the outside edge of the optical disk region, where values range from \numunit{\sigma_{mom2} \sim 22 \text{ to } 35}{\kms}. The PV diagram is presented in the top panel of \figref{fig:pvslice}. It is clear from this PV diagram, as well as the major axis PV diagram in \figref{fig:pvmajor}, that there is low column density gas that spans a wide velocity range. It is probable that this broad low column density component is present throughout the \hi disk of this system but is hidden in the moment 2 map due to much brighter narrow components, particularly in the inner disk region. However, this is not likely the cause of the high moment 2 ring on the edge of the optical disk. 
    
    In \figref{fig:pvslice} we show four line profiles (P1 to P4) extracted at different points along the PV slice. These profiles show that there are multiple distinct overlapping components along the line of sight. This is particularly obvious in P2 and P4.     Given that we already know that there are multiple components making up the overall \hi disk for \gal, it is likely that the multi-component profiles responsible for the high moment 2 ring are as a result of a superposition of the inner and outer disks. This could also explain the offset between the peaks of the different components in P2 and P4.

    \begin{figure*}
        \vspace{-10pt}
        \centering
        \includegraphics[width=\linewidth]{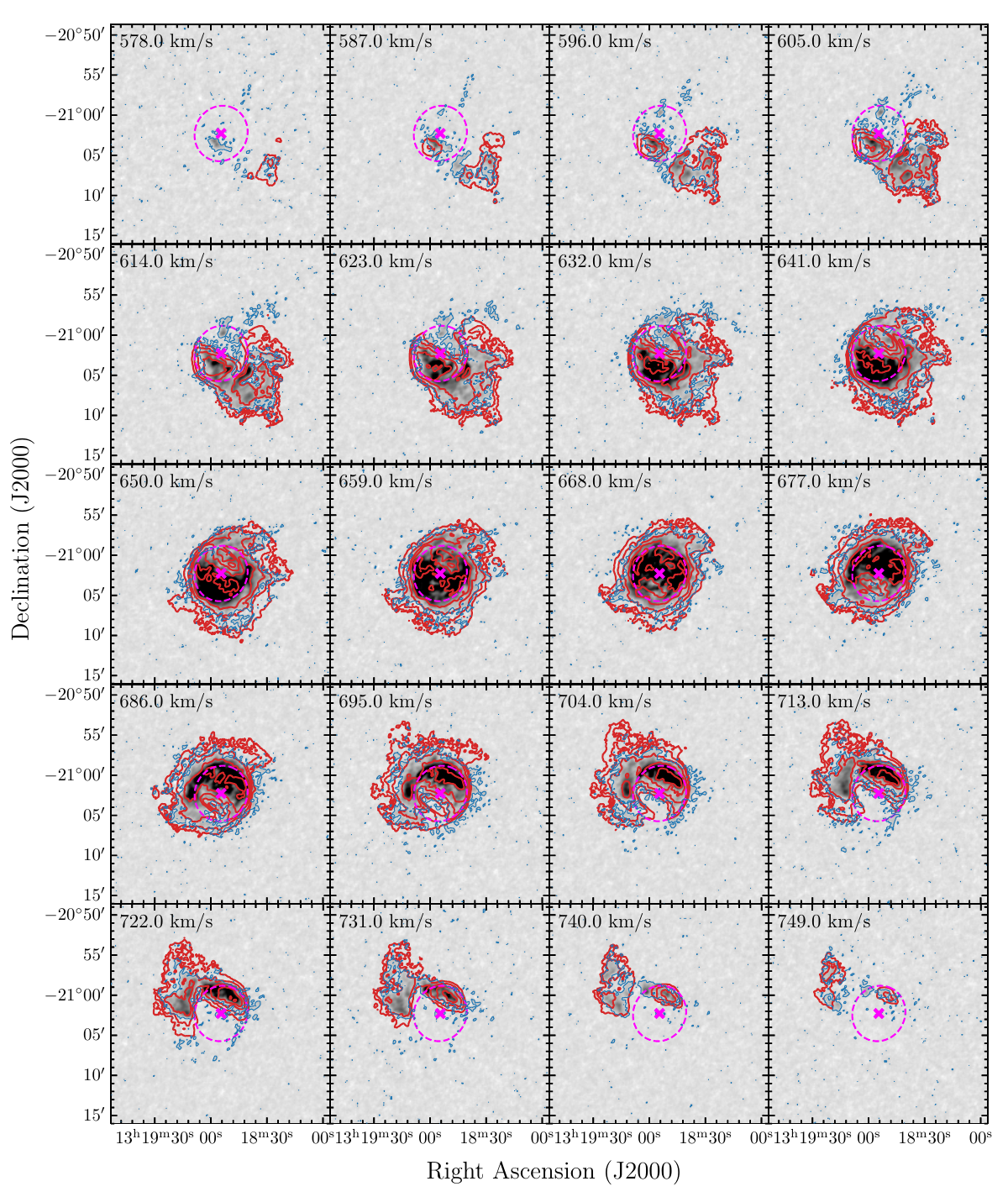}
        \vspace{-15pt}
        \caption{Same as \figref{fig:channelmaps}. The red contours ($0.5\sigma^{data}_{rms} \times 2^n, \, n=0,2,4,..$) represent the inner and outer disk models generated with \texttt{\textsc{Galmod}}.}
        \label{fig:channelmapsmod}
        \vspace{-12pt}
    \end{figure*}

    \begin{figure*}
        \centering
        \includegraphics[width=\linewidth]{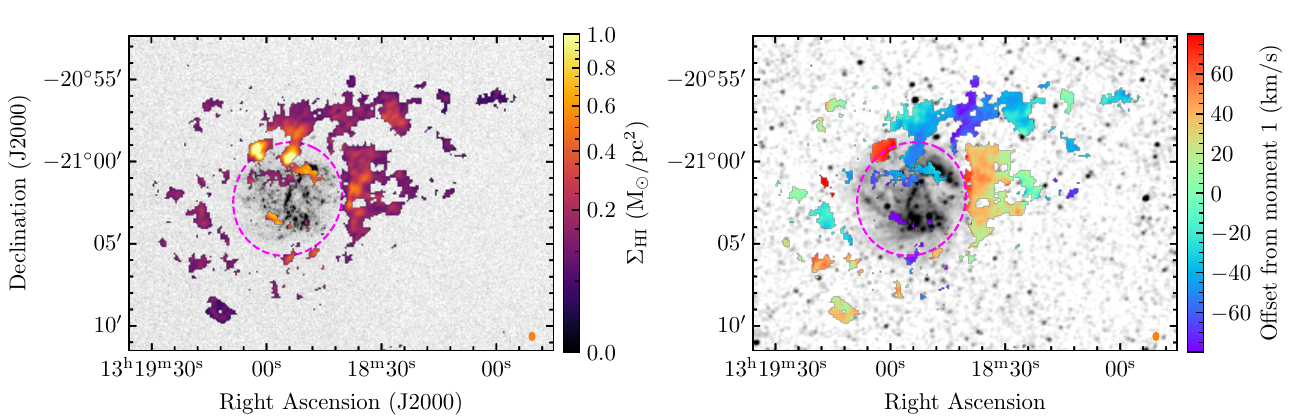}
        \caption{Left: \hi gas surface density of the clumpy gas not contained in the model overlaid on an GALEX FUV image of \gal. Right: the velocity field of the clumpy gas overlaid on the MeerKAT continuum image, the colour shows difference in moment 1 velocity of the clumpy gas and the total velocity field shown in \figref{fig:moment1}.}
        \label{fig:residual}
    \end{figure*}

   \subsection{Modelling the H$\,$\textsc{i} kinematics}
        \label{sec:diskmodel}
         In order to separate out the inner and outer disks from the third component so that we can investigate the origin of the \hi in this system, it is necessary to create a model that describes the morphology and kinematics of the two disks. Modelling the velocity field of \gal using a tilted ring \citep{Rogstad1974} model is not a trivial task, especially considering the degeneracy of parameters such as position angle, inclination, and rotation velocity of almost face-on galaxies such as this one \citep{Jozsa2007a}.

        As discussed in the previous section, there are at least three components to the total \hi disk of this galaxy, however using a version of the \texttt{\sc{Gipsy}} \citep{Allen1985,VanderHulst1992} task \texttt{\sc{Galmod}} as implemented in \textsc{3dBarolo} \citep{DiTeodoro2015}, we have built a simple toy model describing the rotation of the inner and outer disks. For the inner disk, we used the inclination and position angle listed in \tabref{tab:galprop} as measured from near-infrared imaging. The inclination of the outer disk was set to a constant of $53^\circ$ which was estimated from the geometry of the outer \hi contours. 
        
        We experimented with varying the velocity dispersion -- using the average velocity dispersion in the different annuli, as well as different constant values (\numunit{6\text{ to }20}{\kms}), but found it had no real impact on the final model. Thus the velocity dispersion was set to a constant \numunit{15}{\kms} for both the inner and outer disks as this is similar to the median value of the moment 2 map. The position angles for the warped outer disk, and the rotation velocities for both the inner and outer disk were derived from fits of the parameters to the data using the fitting task, \textsc{3dFit}, in \textsc{3dBarolo}. The centre for both the inner and outer disk was set to the optical centre which is consistent with the centre of the \hi.

        The resulting model is shown by the red contours overplotted on the channel maps in \figref{fig:channelmapsmod}. The flux of the model is normalised by that of the data, and is therefore not an independent parameter. The threshold of the lowest red contour in \figref{fig:channelmapsmod} is chosen to match the $3\sigma_{rms}$ (blue) contour of the data. We note that this simple model describes the two disks fairly well apart from the fingers noted in \figref{fig:channelmaps} and the clouds (A,B,C) to the N/NW of the galaxy which we did not include in the model.

        In order to find the emission that is not well described by the model, we blank the data where the model is greater than $0.5\sigma_{rms}$ or greater, this value corresponds to the lowest red contour in \figref{fig:channelmapsmod} which was chosen to match the $3\sigma_{rms}$ data (blue) contour. This gives a ``residual'' data cube from which we can create the usual moment maps of the clumpy gas -- the gas that is not described by either disk model. We impose two further restrictions on what we consider as part of the clumpy gas. The first requirement is to only include high signal-to-noise features in the residual maps, we do this by creating a S/N map using a linewidth of \numunit{20}{\kms}. All pixels in the moment 0 map with S/N$<$5 are excluded. The second requirement is that the values of the corresponding moment 2 map must be \numunit{\mathrm{mom2} < 40}{\kms} -- this is to exclude pixels where the emission could be due to wings of the line not being properly modelled. It would be interesting to see if any of the clumpy gas is related to star formation. We therefore compare the moment maps of the anomalous gas with GALEX FUV observations and radio continuum observations. Both are known to trace emission related to recent or ongoing star formation. {The FUV image is taken from the Survey for Ultraviolet emission in Neutral Gas Galaxies \citep[SUNGG,][]{Wong2016} which collated new and archival targeted observations for a sample of nearby galaxies.} The \hi surface density of the clumpy gas is shown overlaid on a GALEX FUV image of \gal in the left panel of \figref{fig:residual}, and the associated velocity field is overlaid on the \mk continuum image in the right panel of \figref{fig:residual}.

        From both panels of \figref{fig:residual}, we can see that spatially, the clumpy gas is mostly located outside of the main star forming disk, which suggests that this gas is likely not currently involved with any star formation activity. The total mass of this clumpy gas is \numunit{\mathrm{M_{clumpy}} = 2.6 \times 10^7}{\msun} (see \tabref{tab:hicomp}). Since the this gas was not included in the model, it is clearly not rotating with either the inner disk or the outer disk. The velocity field does show a velocity gradient in the different clumps, however it is unclear what the cause of the gradient may be. However it is evident that the clumpy gas on the northern side of the galaxy is responsible for driving the peculiar morphology of the velocity field in that region -- see \figref{fig:moment1}.


\section{Origin of the anomalous gas}
    \label{sec:discuss}

    In the previous section, we have separated out the three components that comprise the \hi of this system: 1--the inner disk that is spatially coincident with the optical disk; 2-- the outer disk which has a kinematic warp, and due to its geometry relative to the inner disk could be considered a polar disk; 3--the clumpy gas that is not described by either disk component. In this section we discuss a number of scenarios that could explain the origin of the peculiar behaviour of the \hi in this system.
    
    \subsection{Interactions with nearby group galaxies}
    
\begin{figure*}
    \centering
    \includegraphics[width=\linewidth]{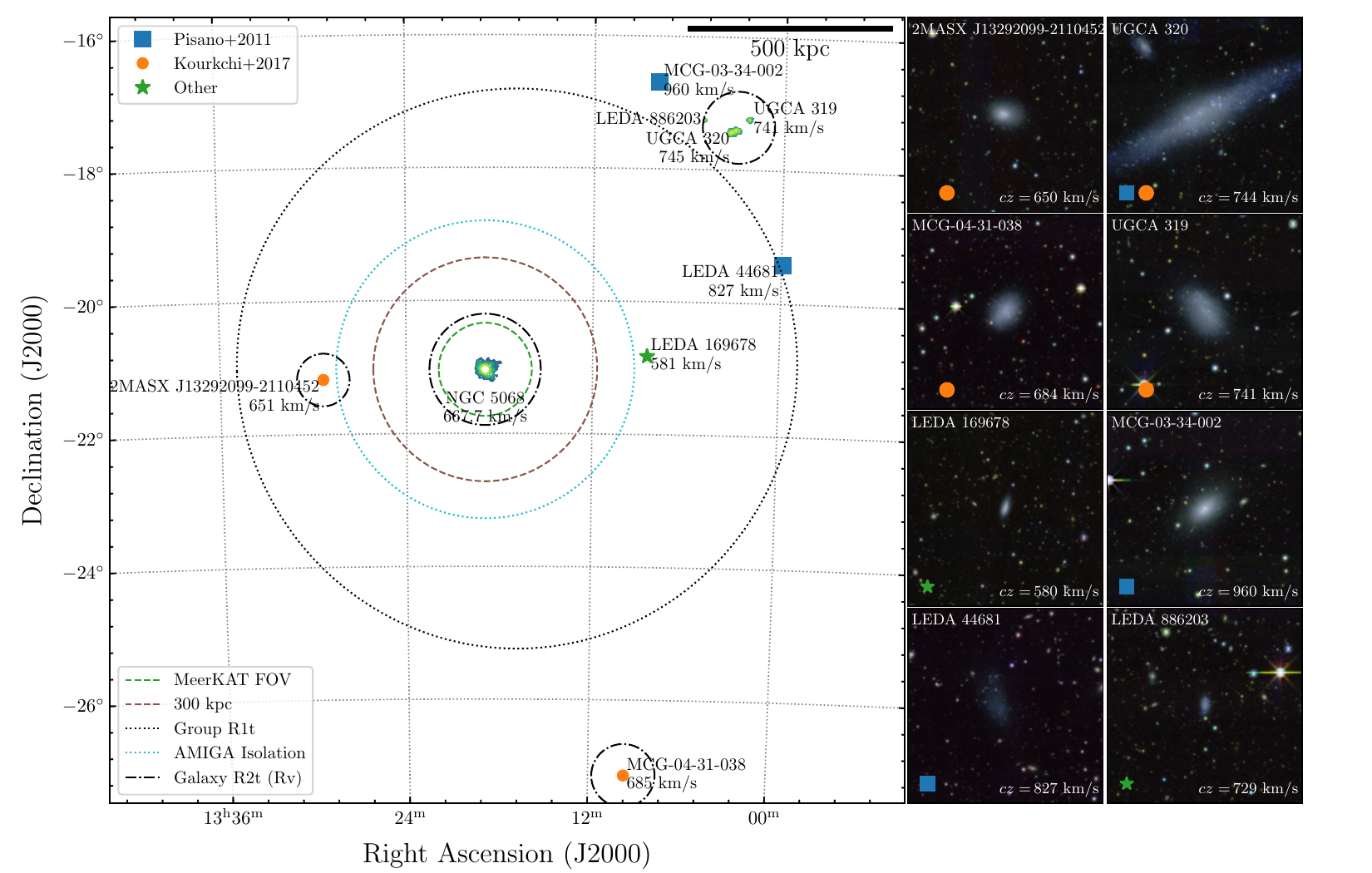}
    \caption{Nearby galaxies around \gal. The galaxies marked with blue squares or orange circles were identified as group members with \gal by \citet{Pisano2011} and \citet{Kourkchi2017}. Left panel: sky distribution of the galaxies, the circles represent regions of different sizes used to determine isolation or identify group members. The R1t and R2t circles are taken from \citet{Kourkchi2017} and are used to group sources together. The AMIGA \citep{Verdes-Montenegro2005a} isolation radius for \gal is represented by the dotted cyan circle. For reference, also plotted is the MeerKAT field of view (dashed green circle), and the brown dashed circle represents a radius of \numunit{300}{\kpc}. The colour $giz$ images on the right are \numunit{5\times 5}{\text{arcmin}^2} cutouts taken from DECaLS DR10.}
    \label{fig:group_isolation}
\end{figure*}

        Figure~\ref{fig:group_isolation} shows the eight spectroscopically confirmed galaxies within a $5^\circ$ radius of \gal. \gal was identified as part of a loose group of galaxies \citep{Pisano2011}, PGC 46400 \citep{Kourkchi2017}. \citet{Karachentsev2017} measured the distances to \gal (\numunit{5.2 \pm 0.2}{\mpc}), LEDA 44681 (\numunit{7.2 \pm 0.3}{\mpc}), UGCA 320 (\numunit{6.03}{\mpc}), and UGCA 319 (\numunit{5.75}{\mpc}) using the redshift-independent tip of the red giant branch method. They conclude that UGCA 320 and UGCA 319 are an isolated pair that are dynamically separate from \gal and LEDA 44681. 
  
         The nearest two galaxies, 2MASX J13292099-2110452 and LEDA 169678, have a projected separation of \numunit{405}{\kpc} and a velocity separation of \numunit{17}{\kms} and \numunit{87}{\kms} respectively. Despite these two close neighbours, \gal is classified as an isolated galaxy using the criteria of \citet{Verdes-Montenegro2005a}: there are no other galaxies with diameters between $1/4$ and 4 times the diameter of the target that lie within a radius that is 20 times the diameter of the potential neighbour. This is based on the assumption that for a galaxy with a diameter of \numunit{25}{\kpc}, an interloper of similar mass moving at a ``field velocity'' of \numunit{150}{\kms} would take \numunit{3}{\text{Gyr}} to close a separation of 20 times the interlopers diameter (\numunit{\sim 500}{\kpc})\footnote{Gas orbital time of \gal is \numunit{\sim 1}{\text{Gyr}}, meaning that any gas would have settled following such an interaction.}.  

        The right panels of \figref{fig:group_isolation} show the composite $giz$ images of the group galaxies (excluding \gal) as observed by DECaLS DR10. The \numunit{5\times 5}{\text{arcmin}^2} colour images show a collection of blue galaxies that are significantly smaller on the sky than \gal which has a semi-major axis of $7.03\arcmin$ (see \tabref{tab:galprop}). Over the redshift range of these galaxies, $1\arcsec$ ranges from \numunit{40}{\text{pc}} to \numunit{57}{\text{pc}}. Based on the optical colours, sizes, and morphologies of this collection of galaxies, it is not likely even the closest two neighbours discussed above have ever interacted with \gal.

    \subsection{Stripped gas from a passing neighbour}
    
    What if the anomalous gas is the result of a gas rich unidentified neighbour passing by \gal and losing all of its gas to \gal? In such scenario the gas may have been stripped off the neighbour which continued on past \gal, and part of the gas settled into the potential of \gal as the outer disk with the remainder of the gas remaining as clumps beyond the outer disk. We do not believe this to be a likely origin of the anomalous gas. Below we discuss why the scenario is improbable using simple assumptions based on the mass of the \hi in the outer disk.
    
    If the interaction had taken place, it would have been more than \numunit{1}{\text{Gyr}} in the past as the outer gas has already completed at least one orbit as it has settled into a disk-like component within the potential of \gal.

    {Given the \hi mass of the outer gas (\numunit{\mhi = 8.9\times 10^7}{\msun}), the stellar mass of the neighbour would be roughly \numunit{4.45\times 10^7}{\msun}, this assumes $\mhi/\mstar = 2$, which is consistent with the Small Magellanic Cloud \citep{Stanimirovic1999} and other dwarf galaxies \citep{Huang2012}. The stellar mass ratio between the two galaxies is $1:50$, which means that the neighbour would not likely leave an imprint on \gal as a result of the interaction.}

    {Using the \mhi--$D_\hi$ relation from \citet[][Eq. 2]{Wang2016}, and assuming the diameter of the optical disk is $0.8\times D_\hi$, we determine the neighbour is roughly \numunit{D_{opt} \sim 4}{\kpc} ($\sim 1.4\arcmin$ at \numunit{5.2}{\mpc}). With a conservative stellar mass to light ratio of 0.6, we calculate the mean $r$-band surface brightness of the neighbour to be \numunit{\mu_r = 23.7}{\text{mag arcsec}^{-2}}. If this neighbour passed by \gal with a maximum velocity of \numunit{v = 100}{\kms}, it would be located within a \numunit{100}{\kpc} ($36\arcmin$) radius after \numunit{1}{\text{Gyr}}. The $5\sigma$ detection threshold in the DECaLS $r$-band image is \numunit{\mu_r = 26.3}{\text{mag arcsec}^{-2}} which means that the neighbour would be visible in the DECaLS data, however there is no likely candidate in the DECaLS imaging.}

    \subsection{Minor merger with a gas rich galaxy}
    
    Since there is no neighbour in our search area, what if \gal underwent a merger with a gas-rich low mass satellite? Given the \hi mass difference between the inner and outer disks, this would suggest a merging ratio of 1:10. Such an event would have to have taken place more than \numunit{1}{\text{Gyr}} ago as the outer \hi has settled in the potential of \gal. {Minor mergers are usually defined as where the lower mass galaxy is more than four times less massive than the larger galaxy. Some studies \citep[][]{Lotz2008,Conselice2005} have suggested that the timescales for minor mergers can vary from \numunit{\tau\sim 0.3\text{ to } 1}{\text{Gyr}} assuming an initial separation of \numunit{30}{\kpc}. More recently, \citet{Conselice2022} have shown using hydronamical simulations that the timescale of mergers is also dependent on both redshift and mass ratio of the merging galaxies. Based on the \citet{Conselice2022} study, the merger timescale for a $z\sim 0$ merger with a mass ratio of 1:10, could range between \numunit{\tau \sim 1.7\text{ to }2}{\text{Gyr}}, however this assumes that the galaxies have \numunit{\mstar > 10^9}{\msun}. Given that the \hi in the \gal system has already settled, it is likely that if a minor merger had occurred it would have happened more than \numunit{2}{\text{Gyr}} ago. Despite the mass difference between the constituent galaxies, minor mergers are known to} have an impact on the morphology and kinematics of the larger galaxy \citep[e.g.][]{Kazantzidis2009,Qu2011,Martin2018,Ghosh2022}. 

     The optical morphology of \gal is undisturbed, and does not appear to have undergone any recent interactions. However it is possible that in such a scenario the main stellar disk of \gal was undisturbed. Mergers are thought to be one of the origin scenarios of polar disks \citep[e.g.][and references therein]{Stanonik2009}. While there is no clearly observable stellar component within the available deep optical imaging from DECaLS associated with the outer disk, it is possible that the outer disk could be caused by a minor merger and the associated stellar disk has been stretched out below detection limits. This scenario, however, does not explain the clouds which are a largely located on the northern side of the galaxy. 
     
     If all the anomalous gas (all \hi gas not associated with the inner disk) from \gal could be traced back to one event such as a minor merger, given the time that has likely elapsed (\numunit{>2}{\text{Gyr}}), one would expect that the clouds would have settled into a more symmetric distribution around the galaxy. While we cannot conclusively rule out a minor merger as a possible origin scenario given that this scenario does not neatly explain all the gas, we do not do assign this scenario a high probability of being the origin of the anomalous gas.

   \subsection{Fountain triggered accretion}

    Galactic fountains can be an important mechanism by which gas is recycled within a galaxy system \citep[see, for example,][]{Shapiro1976,Fraternali2006}: gas is expelled from the main disk by processes associated with star formation such as supernovae and solar winds. {The expelled gas} mixes with the hotter gas in the halo and then the mixed gas cools and condenses before it is re-accreted onto the galaxy. This fountain driven accretion of gas means that the accreted gas contains both the original expelled gas, and gas from the halo, the net flow of the gas is into the galaxy. Given that this process arises as a result of activity in the stellar disk, the accreted gas is likely to be coincident with the stellar disk.

    Simulations suggest that gas accreted onto the galaxy subsequent to mixing with gas ejected from the galaxy disk through fountain processes are likely to have a higher metal content than primordial gas \citep[see review by][]{Almeida2014,Brook2014,Tumlinson2017} which is thought to occur along filaments into the galaxy. Thus looking at the gas phase metallicity of the star forming regions on the edge of the galaxy disk, particularly where the anomalous gas and the star forming regions overlap, may provide some insight into the origin of gas \citep{Howk2018}. 
    
    \gal has been observed by a number of multi-wavelength surveys, but two surveys in particular have looked at the gas-phase metallicity of the star forming regions in \gal: PHANGS-MUSE \citep{Emsellem2022,Williams2022} and TYPHOON \citep[M. Seibert et al. in prep;][]{Grasha2022}. Both \citet{Williams2022} and \citet{Grasha2022} show that the metallicity of the gas in \gal cannot be explained by the simple radial model that describes the negative metallicity gradient seen in star-forming galaxies. \citet[][Fig.~3]{Grasha2022} in particular show that the H$\,$\textsc{ii} regions on the northern side of the galaxy are more metal poor than regions at the same radius on the southern side of the galaxy. 

    Given the distribution of the star forming regions throughout \gal (see the background FUV and continuum images in \figref{fig:residual}), if fountain-driven accretion was responsible for the anomalous gas in \gal, it is plausible that the distribution of the clumpy gas not explained by the disks would be more equally distributed around the galaxy. We conclude that while it is very likely that there are fountain-processes ongoing in \gal, it is not the likely source of the anomalous \hi gas.

    \subsection{Accretion of gas along a filament}
    
    The final scenario that we consider is that the anomalous gas is a result of gas accretion external to the galaxy halo. How this accretion occurs, and what the observational signatures of such accretion look like are still unclear. Simulations have shown that in galaxies the size of \gal, accretion is expected to occur along filaments, condensing into clumps and clouds as it nears the galaxy disk \citep{Kaufmann2006,VandeVoort2011b, Wetzel2015, Cornuault2018, Iza2022}.
 
    Several observational signatures have been suggested as proof of these accretion modes. Due to the gas condensing as it reaches the \hi disk, it is expected that the gas is accreted in clouds or clumps. Clouds or clumps with anomalous velocities have been observed around the Milky Way, these clouds, are known as intermediate (IVC) and high (HVC) velocity clouds \citep[see review by][and references therein]{Wakker1997,Wakker1999}. {The clouds around the Milky Way have been shown to }trace the flow of accreting gas, both as a result of extragalactic accretion \citep{Wakker2007,Wakker2008, Peek2008}, and re-accretion of gas relating to a galactic fountain \citep{McClure-Griffiths2013,Marasco2022}. 
    
    The left panel of \figref{fig:residual} shows that the anomalous gas associated with \gal is clumpy, it should be noted that majority of this clumpy gas is very low column density (\numunit{\nhi < 10^{19}}{\cmt}). As discussed in Section~\ref{sec:globalprop}, the clouds on the north western side of the galaxy range in \hi mass from \numunit{\mhi \sim 10^4}{\msun} to \numunit{\mhi \sim 10^5}{\msun} which is consistent with the \hi masses of the IVC/HVC complexes around the Galaxy \citep{Wakker2008}. The velocity field of the anomalous gas relative to the galaxy velocity field (see \figref{fig:moment1}) is shown in the right panel of \figref{fig:residual}. The velocities of the anomalous gas show that the clouds coincident with the star-forming disk deviate from the overall rotation of the main \hi disk. 
    
    In Section 5.3, we mentioned how mergers are one of the formation scenarios of polar disks. More recently, it has been shown that accretion could also trigger the formation of a polar disk, and this case an associated stellar component would not be expected \citep{Maccio2006,Brook2008, Stanonik2009}. Another feature present in polar disks when accretion is considered the origin, is kinematic warps \citep[e.g.][]{Brook2008}. While the origin of the warps are not yet clear, it has been shown that they can be explained as the result of skewed angular momentum of gas accreting due to cosmic infall \citep{Ostriker1989,Jiang1999,Sanchez-Salcedo2006,Rand2008}. 

    Both the morphology and the kinematics of the anomalous gas in combination of the warped (possibly polar) outer disk of \gal point to ongoing accretion. It is interesting to note to that some of the clumpy gas is coincident with the optical disk right where there is some of the brightest star formation on the northern ridge of the galaxy. These star forming regions also have some of the lowest gas phase {metallicity} in the maps presented in \citet{Grasha2022}.

    Putting all of this together, we propose the following scenario: gas is being accreted along a filament that is aligned with the north/north western edge of the galaxy. This gas condenses into the clumps/clouds when it gets close to the galaxy disk, over time this accretion has given rise to what is now observable as the outer disk. Within this scenario, it is possible that not all the clouds feed the outer disk, but some are pulled into the star-forming disk enhancing the the star formation on the northern side of galaxy.


\section{Summary}
In this paper we have presented new MeerKAT observations of the \hi in \gal taken as part of the MeerKAT Large Survey Project, MHONGOOSE. The combination of the impressive sensitivity and resolution of these data have revealed a number of interesting features in the \hi disk of \gal. We have identified three separate components to the total \hi disk: the inner disk, the outer disk which shows signatures of a kinematic warp and given the geometry could be a polar disk, and lastly the clumpy region to the north western side of the galaxy. We constructed a model that contained a regularly rotating inner disk plus a more inclined warped outer disk that described the majority of the gas kinematics adequately. Using the model, we were able to isolate a significant amount of clumpy gas that was not well explained by the model of the disks. This clumpy gas accounts for $\sim 2\%$ of the total \hi mass of the system and also contains the clumps to the north/north west of the galaxy.

We explored a number of different possible origin scenarios for the anomalous gas:
\begin{itemize}
\item[(a)] Interaction with the nearby group galaxies -- the projected distances between the galaxies make this an unlikely scenario.
\item[(b)] Stripped gas from a passing neighbour -- assuming the neighbour survived the passage, we find no evidence for its existence.
\item[(c)] A minor merger with a gas rich neighbour -- we cannot conclusively exclude this scenario, but do not believe it is the most likely scenario as it does not explain the presence of the clumpy gas on only one side of the galaxy.
\item[(d)] Fountain triggered accretion -- accretion is likely, but this scenario would suggest a more metal enrichment than what is observed in the outer star forming regions, and for the clumpy gas distribution to follow the locations star forming regions.

\item[(e)] Accretion of gas along a filament -- this scenario neatly explains the kinematics and morphology of all the anomalous \hi.
\end{itemize}

While the last scenario is the most likely based on the data we have presented in this paper, further detailed modelling of the gas is needed to understand if and how the gas is transported between the inner and outer disks. Unambiguous detection of cold mode accretion of gas onto galaxies will require the detection of \hi gas at column densities of \numunit{\nhi \sim 5 \times 10^{18}}{\cmt} or lower, and at cloud scale resolution in combination of other tracers of cold gas such as Lyman-$\alpha$ \citep{VandeVoort2012, Ao2020, Kacprzak2017}. Simulations and some recent observations of high redshift galaxies suggest that Lyman-$\alpha$ emission traces cold gas being accreted onto the galaxies \citep[e.g.][]{Dijkstra2009,Daddi2021}. However, with the currently available facilities, the ongoing MHONGOOSE survey provides the best chance of detecting and studying in detail, the low column density \hi that could be linked to cold gas accretion.

\begin{acknowledgements}
    We acknowledge useful discussions on the interpretation of this data with members of the MHONGOOSE team. JH thanks Nathan Deg for helpful discussions on polar ring galaxies. We thank the MeerLICHT team for their help with the MeerLICHT imaging, in particular: Paul Vreeswijk, Danielle Pieterse, Steven Bloemen, Paul Groot, and Patrick Woudt. Thanks to Filippo Fraternali and Thijs van der Hulst for useful discussions about the interpretation of the \hi in this galaxy.

    This project has received funding from the European Research Council (ERC) under the European Union’s Horizon 2020 research and innovation programme grant agreement no. 882793, project name MeerGas. 

    PK acknowledges financial support by the German Federal Ministry of Education and Research (BMBF) Verbundforschung grant 05A20PC4 (Verbundprojekt D-MeerKAT-II).
    
    DJP and NZ are supported through the South African Research Chairs Initiative of the Department of Science and Technology and National Research Foundation.
    
    KS acknowledges support from the Natural Sciences and Engineering Research Council of Canada (NSERC).

    BKG acknowledges the financial support of the European Union’s Horizon 2020 Research and Innovation Programme (ChETEC-INFRA -- Project no. 101008324).

    LVM acknowledges financial support from the grant CEX2021-001131-S funded by MCIN/AEI/ 10.13039/501100011033, from the grant PID2021-123930OB-C21 funded by MCIN/AEI/10.13039/501100011033, by ``ERDF A way of making Europe'' and by the European Union.

    LC acknowledges the financial support from the Chilean Agencia Nacional de Investigaci\'{o}n y Desarrollo through the grant Fondecyt Regular 1210992.

    This paper makes use of MeerKAT data. The MeerKAT telescope is operated by the South African Radio Astronomy Observatory, which is a facility of the National Research Foundation, an agency of the Department of Science and Innovation. 

    This research made use of Astropy,\footnote{http://www.astropy.org} a community-developed core Python package for Astronomy \citep{Astropy2013, Astropy2018}.

    Part of the data published here have been reduced using the CARACal pipeline, partially supported by ERC Starting grant number 679627 ``FORNAX'', MAECI Grant Number ZA18GR02, DST-NRF Grant Number 113121 as part of the ISARP Joint Research Scheme, and BMBF project 05A17PC2 for D-MeerKAT. Information about CARACal can be obtained online under the URL: \url{https://caracal.readthedocs.io}.
      
\end{acknowledgements}

\bibliographystyle{aa}
\bibliography{references}

\end{document}